\documentclass[preprint,10pt]{elsarticle}

\usepackage{graphicx}

\usepackage{amssymb}

\newcounter{bla}

\journal{Computer Physics Communications}

\begin{document}

\begin{frontmatter}



\title{GR@PPA 2.8: initial-state jet matching for weak boson production 
processes at hadron collisions}


\author{Shigeru Odaka\corref{odaka}}
\author{Yoshimasa Kurihara}

\cortext[odaka] {Corresponding author.\\
\textit{E-mail address:} shigeru.odaka@kek.jp}

\address{High Energy Accelerator Research Organization (KEK)\\
1-1 Oho, Tsukuba, Ibaraki 305-0801, Japan}

\begin{abstract}
The initial-state jet matching method introduced in our previous studies 
has been applied to the event generation of single $W$ and $Z$ production 
processes and diboson ($W^{+}W^{-}$, $WZ$ and $ZZ$) production processes 
at hadron collisions in the framework of the GR@PPA event generator. 
The generated events reproduce the transverse momentum spectra of 
weak bosons continuously in the entire kinematical region.
The matrix elements (ME) for hard interactions are still at the tree level.
As in previous versions, the decays of weak bosons are included in the 
matrix elements.
Therefore, spin correlations and phase-space effects in the decay of weak 
bosons are exact at the tree level.
The program package includes custom-made parton shower programs as well as 
ME-based hard interaction generators 
in order to achieve self-consistent jet matching.
The generated events can be passed to general-purpose event generators 
to make the simulation proceed down to the hadron level.
\end{abstract}

\begin{keyword}
GRACE; Hadron collision; Event generator; Jet matching; Parton shower; 
Weak boson production
\end{keyword}

\end{frontmatter}



\noindent
{\bf PROGRAM SUMMARY}

\begin{small}

\noindent
{\em Manuscript Title:}  GR@PPA 2.8: initial-state jet matching 
for weak boson production processes at hadron collisions\\
{\em Authors:} S. Odaka, Y. Kurihara\\
{\em Program Title:} GR@PPA 2.8\\
{\em Journal Reference:}                                      \\
{\em Catalogue identifier:}                                   \\
{\em Licensing provisions:} none\\
{\em Programming language:} Fortran; 
with some included libraries coded in C and C++\\
{\em Computer:} all\\
{\em Operating system:} any UNIX-like system\\
{\em RAM:} 1.6 Maga bytes at minimum\\
{\em Supplementary material:}                                 \\
{\em Keywords:} GRACE; Hadron collision; Event generator; 
Jet matching; Parton shower; Weak boson production\\
{\em Classification:} 11.2\\
{\em External routines/libraries:} bash and Perl for the setup, 
and CERNLIB, ROOT, LHAPDF, PYTHIA according to the choice of users\\
{\em Does the new version supersede the previous version?:}
No, this version supports only a part of the processes included 
in previous versions.\\
{\em Nature of problem:}
We need to combine those processes including 0 jet and 1 jet 
in the matrix elements using an appropriate matching method, 
in order to simulate weak-boson production processes 
in the entire kinematical region.
\\
{\em Solution method:} 
The leading logarithmic components to be included in parton distribution 
functions and parton showers are subtracted from 1-jet matrix elements.
Custom-made parton shower programs are provided to ensure satisfactory 
performance of the matching method.\\
{\em Reasons for the new version:}
An initial-state jet matching method has been implemented.\\
{\em Summary of revisions:}
Weak-boson production processes associated with 0 jet and 1 jet can be 
consistently merged using the matching method.\\
{\em Restrictions:} The built-in parton showers are not compatible with 
the PYTHIA new PS and the HERWIG PS.\\
{\em Unusual features:} A large number of particles may be produced 
by the parton showers and passed to general-purpose event generators.\\
{\em Running time:} About 10 min. for initialization plus 25 sec. for 
every 1k-event generation for $W$ production in the LHC condition, 
on a 3.0-GHz Intel Xeon processor with the default setting.\\

\end{small}

\section{Introduction}

The inclusive production cross section of a final state, $A$, from collisions 
of two hadrons, $h_{1}$ and $h_{2}$, at a squared center-of-mass (cm) energy 
of $s$ is usually evaluated as
\begin{eqnarray}
  \sigma_{h_{1}h_{2} \rightarrow A+X} = \sum_{a,b,i} \int_{0}^{1} dx_{1} 
  \int_{0}^{1} dx_{2} \int d{\hat \Phi_{A_{i}}}
  f_{h_{1} \rightarrow a}(x_{1},\mu_{F}^{2})
  f_{h_{2} \rightarrow b}(x_{2},\mu_{F}^{2})
  { d{\hat \sigma}_{ab \rightarrow A_{i}}({\hat s}) \over d{\hat \Phi_{A_{i}}}}
  \delta({\hat s} - x_{1}x_{2}s),
\label{eq:xsec}
\end{eqnarray}
where $f_{h_{k} \rightarrow a}(x_{k},\mu_{F}^{2})$ is the parton distribution 
function (PDF) representing the existence probability of the parton $a$ 
(a light quark or a gluon) inside the hadron $h_{k}$ with a momentum fraction 
of $x_{k}$ at a certain energy scale of $\mu_{F}$ (factorization scale). 
The factor 
$d{\hat \sigma}_{ab \rightarrow A_{i}}({\hat s})/d{\hat \Phi_{A_{i}}}$ 
represents the differential cross section of the hard interaction 
that produces the final state $A_{i}$ from the collision of two partons, 
$a$ and $b$, perturbatively calculated according to fixed-order matrix 
elements (ME) at a squared cm energy of ${\hat s}$.
The final state $A$ may consist of several sub-states $A_{i}$ 
at the parton level; for instance, "jet" production includes a variety 
of light quark and gluon productions.
Such a generalization is necessary in hadron collisions 
because it is difficult to separate the sub-states experimentally.

Monte Carlo (MC) event generators are, in principle, an MC integration 
of Eq.~(\ref{eq:xsec}), 
in which the distribution of the sampling points is controlled so that 
the frequency is proportional to the differential cross section. 
Thus, each sampling point (event) can be treated just like an event produced 
by actual interactions.
MC event generators are mainly used for evaluating 
the detection efficiency and acceptance of experiments; 
therefore, they are desired to exclusively reproduce 
the actual phenomena as precisely as possible.
The initial-state QCD activities reproduced by PDFs in Eq.~(\ref{eq:xsec}) 
not only alter the longitudinal momentum distribution of constituent partons, 
but also produce transverse activities 
which result in a finite transverse recoil of the hard interaction system 
and additional hadronic activities visible in detectors.
Experimentalists want MC event generators to reproduce such transverse 
activities as well, 
since all these effects may alter the performance of detectors. 

Parton showers have been developed to fulfill the above requirement;
they reproduce the transverse activities of QCD radiations 
as well as the resultant longitudinal evolution evaluated in PDFs. 
Parton showers are a recursive solution of 
the DGLAP equation~\cite{Gribov:1972ri,Altarelli:1977zs,Dokshitzer:1977sg} 
from which the perturbative part of PDFs is derived.
Thus, in principle, the PDFs in Eq.~(\ref{eq:xsec}) can be replaced with 
parton showers, except for the non-perturbative components of the PDFs.
However, in most of the MC event generators currently available, 
PDFs are directly employed for evaluating the initial-state conditions 
of the hard interactions, and parton showers are adopted only as models 
for simulating the initial-state QCD activities 
by using a technique of the "backward evolution"~\cite{Sjostrand:1985xi}.
The identity between the PDF and PS is not seriously considered 
in such simulations.

PDFs sum up the collinear QCD corrections, which result in large logarithmic 
terms, to all orders of the coupling constant $\alpha_{s}$ 
in order to improve the convergence of the perturbative calculations 
for hard interactions.
The summation is limited by an arbitrary energy scale (factorization scale), 
and the resultant cross section depends on the choice of this parameter.
The factorization scale, $\mu_{F}^{2}$, is defined as the upper limit on the 
momentum transfer, $Q^{2} = -t$, of the radiations considered in the summation. 
This scale is usually taken to be equal to a typical energy scale, 
such as ${\hat s}$ or $|{\hat t}|$, of the considered hard interaction 
because non-collinear components missing in PDFs become significant 
around this scale.
Parton showers should reproduce the perturbative part of PDFs; 
hence, if we consider the identity between PDF and PS seriously, 
$Q^{2}$ of PS branches must also be limited by $\mu_{F}^{2}$.
As a result, the transverse recoil of the system $A$ in Eq.~(\ref{eq:xsec}) 
is limited.
Since there is no such limitation in actual phenomena, 
the inclusion of harder radiations is necessary to reproduce the phenomena 
in the entire phase space.

In order to add the hard radiation effect to Eq.~(\ref{eq:xsec}), 
we have to anyhow include the effects of interactions 
that produce at least one additional parton in association with the production 
of $A$ ($A$ + 1-jet process).
A straightforward approach is to replace the hard interaction part 
in Eq.~(\ref{eq:xsec}) with that for the $A$ + 1-jet process, and then, 
to add the resultant cross section or an event sample to the results 
for the production of $A$ ($A$ + 0-jet process).
Equation~(\ref{eq:xsec}) does not have any problem if the final state $A$ 
is a color-singlet state, for instance, a single weak boson.
However, we encounter serious problems 
if we replace it with $A$ + 1-jet processes. 

Partons in the final state (jets) are in most cases produced 
as a result of QCD interactions.
Double counting may occur because PDFs and parton showers are also 
consequences of QCD interactions. 
We encounter this problem even when we simply try to generate those events 
including partons in the final state of the hard interaction. 
The same problem also occurs when we try to combine processes 
having different jet multiplicities, 
for instance, the $A$ + 0-jet process and the $A$ + 1-jet process.
The simplest solution to this problem is to clearly separate the phase space 
for the radiations using $\mu_{F}$. 
Since radiation effects at $Q^{2} < \mu_{F}^{2}$ are taken into account 
in PDFs and parton showers, 
the hard interaction part should include only those radiations having 
$Q^{2} > \mu_{F}^{2}$. 
Although double counting can be avoided with this solution, 
another serious problem (mismatch) may occur owing to the existence 
of non-collinear contributions disregarded in PDFs and parton showers, 
which may become significant around $\mu_{F}^{2}$. 

Several solutions to the double counting problem have been proposed 
and implemented in MC event generators,
such as the ME correction in PYTHIA~\cite{Miu:1998ju} and 
HERWIG~\cite{Seymour:1994df}, 
and the CKKW method~\cite{Catani:2001cc} implemented 
in Sherpa~\cite{Gleisberg:2008ta}. 
The MLM prescription in AlpGen~\cite{Mangano:2002ea} can be considered 
as an alternative to the CKKW method.
These are solutions for leading-order (LO) event generators.
For complete next-to-leading order (NLO) event generation, 
a subtraction method is applied in MC@NLO~\cite{Frixione:2002ik,Frixione:2008ym} 
and a suppression method is used in POWHEG~\cite{Nason:2004rx,Alioli:2010xd}.

GR@PPA (GRace At Proton-Proton/Antiproton)\footnote{
{\tt http://atlas.kek.jp/physics/nlo-wg/grappa.html}.} 
is a Monte Carlo event generator package for simulating interactions 
at proton-proton and proton-antiproton collider experiments.
It is an extension of the GRACE system \cite{Ishikawa:1993qr,Yuasa:1999rg} 
to hadron collision interactions.
GRACE is a powerful tool for deriving the differential cross section 
of hard interactions at the parton level, 
$d{\hat \sigma}_{ab \rightarrow A_{i}}({\hat s})/d{\hat \Phi_{A_{i}}}$ 
in Eq.~(\ref{eq:xsec}), and for generating events according to it 
with the help of BASES/SPRING \cite{Kawabata:1985yt,Kawabata:1995th}.
GR@PPA provides a mechanism for adding the effects of the initial-state 
variation in the flavor and momentum according to PDF 
and for achieving the generalization of the final state, 
as described symbolically in Eq.~(\ref{eq:xsec}).
The previous releases of GR@PPA~\cite{Tsuno:2002ce,Tsuno:2006cu} 
include many multi-body (multi-jet) production processes, 
such as $W$ and $Z$ + jets, diboson ($W^{+}W^{-}$, $WZ$, $ZZ$) + jets, 
top-pair + jet, and QCD multi-jets.
However, its application is restricted because the jet matching discussed above 
is not taken into consideration.

As described in previous reports, we have proposed a solution 
(matching method)~\cite{Kurihara:2002ne} to the double counting problem, 
and we have shown its feasibility in $W$ boson production~\cite{Odaka:2007gu}.
We have also shown that if we apply the method to $Z$ boson production 
in the GR@PPA event generator, the generated events reproduce 
the $p_{T}$ spectrum of $Z$ bosons measured at Fermilab Tevatron 
with surprisingly high precision over the entire measurement 
range~\cite{Odaka:2009qf}.
In this report, we describe a new version of the GR@PPA event generator 
package, GR@PPA 2.8, in which our matching method is applied 
not only to the single $W$ and $Z$ boson production processes 
but also to the diboson ($W^{+}W^{-}$, $WZ$ and $ZZ$) production processes 
in proton-proton and proton-antiproton collisions. 

Although our matching method is designed with the objective of developing 
NLO event generators, 
where the inclusion of one additional parton in the final state is necessary, 
the event generation in GR@PPA 2.8 is currently at the tree level
because virtual corrections are yet to be included.
Jet matching is accomplished by subtraction, as in the case of MC@NLO.
However, subtraction is carried out in a limited phase space in our method, 
whereas there is no such limitation in MC@NLO.
The limited application of subtraction can potentially enable us 
to easily extend the matching method to the final state. 
If the extension is realized successfully, 
it can be used for the matching between multi-jet production processes, 
similar to the CKKW method. 
Because our method is based on the concept 
that higher jet multiplicity generators are used to supplement the deficits 
in lower jet multiplicity generators, 
the necessity for higher jet multiplicity processes will not be as critical 
as that in the CKKW method.

Although the process-independent framework and process-dependent packages 
were provided separately in the previous version~\cite{Tsuno:2006cu}, 
the current version provides an all-in-one package because our matching method 
can be presently applied only to a subset of the processes implemented 
in the previous version. 
Though other process packages in the previous version can be imported 
in principle, 
such a modification is not recommended because careful treatments must be 
required for the execution, and the performance has not been tested.
The features supported in previous versions~\cite{Tsuno:2002ce,Tsuno:2006cu}, 
such as inclusion of weak-boson decays in matrix elements, 
finite decay widths of weak bosons, 
branching ratios of weak bosons tuned to measurement data, 
inclusion of CKM non-diagonal couplings, 
and $Z$-photon mixing,
are also supported in the current version.

The remainder of this report is organized as follows. 
The matching method applied in GR@PPA 2.8 is described 
in Section~\ref{sec:method}.
In general, the detailed implementation of the solution to the double counting 
problem depends on the parton shower used in the event generation.
We provide our own parton showers in this package in order to ensure 
satisfactory performance of our method.
Parton showers are provided for the initial state and the final state.
The former is crucial for our matching method, 
whereas the latter remains experimental and is implemented for completeness.
These parton showers are described in Section~\ref{sec:ps}.
Section~\ref{sec:run} contains instructions for the installation 
of libraries and execution of sample programs.
Some results that have not been presented in previous reports are presented 
in Section~\ref{sec:res}. 
Practical performance parameters such as the program size and CPU time 
are presented in Section~\ref{sec:perf}. 
Finally, a summary is provided in Section~\ref{sec:sum}.
 
\section{Initial-state jet matching}
\label{sec:method}

The concept of our matching method has been described in a previous 
report~\cite{Kurihara:2002ne}.
Hereafter, we assume that $A$ represents a single weak boson 
or a diboson system.
Our concept is as follows. 
We preserve the event generation according to Eq.~(\ref{eq:xsec}),
with the factorization scale $\mu_{F}$ chosen as usual.
Then, we separately generate events simulating the radiation 
contributions that are missing in Eq.~(\ref{eq:xsec}), 
non-collinear contributions and larger $Q^{2}$ ($> \mu_{F}^{2}$) 
contributions, using the matrix elements for the $A$ + 1-jet process.
If these two event samples are combined, 
we should obtain an event sample that covers all the radiation effects 
that may affect the production kinematics of the system $A$.
Since the $A$ + 1-jet process is of the first order in QCD, 
it contains only the leading-order contribution of parton showers.
This contribution can be factorized and can be evaluated 
using the $A$ + 0-jet matrix element and a radiation factor.
Therefore, we can derive the desired $A$ + 1-jet process by subtraction.
We call this method the limited leading-log (LLL) subtraction.

The LLL subtraction is carried out at the matrix-element (ME) level as
\begin{eqnarray}
  \left|\mathcal{M}_{A+1}^{(sub)}({\hat s}_{A+1}, {\hat \Phi}_{A+1}; 
  \mu_{R})\right|^{2} 
  &=& \left|\mathcal{M}_{A+1}({\hat s}_{A+1}, {\hat \Phi}_{A+1}; 
  \mu_{R})\right|^{2}  \nonumber\\
  &&- \sum_{i}\left|\mathcal{M}_{A }({\hat s}_{A}, {\hat \Phi}_{A,i}; 
  \mu_{R})\right|^{2}
  f_{LL, i}(Q_{i}^{2},z) \theta(\mu_{F}^{2} - Q_{i}^{2}),
\label{eq:sub}
\end{eqnarray}
where the first term on the right-hand side represents the exact ME 
for $A$ + 1-jet production at a squared cm energy of ${\hat s}_{A+1}$.
The factor $f_{LL,i}(Q_{i}^{2},z)$ is the radiation factor 
in the leading-log approximation, 
and $|\mathcal{M}_{A }({\hat s}_{A}, {\hat \Phi}_{A,i}; \mu_{R})|^{2}$ 
represents the ME for the non-radiative ($A$ + 0-jet) subsystem 
having a squared cm energy of ${\hat s}_{A} = z{\hat s}_{A+1}$. 
The Lorentz boost and angular rotation of the final state $A$, 
owing to the jet radiation, 
is taken into account in the calculation of the $A$ + 0-jet ME.
The matrix elements are evaluated using the first-order strong coupling, 
expressed as
\begin{equation}\label{eq:alpha_s}
	\alpha_{s}(Q^{2}) = {4 \pi \over \beta_{0} \ln(Q^{2}/\Lambda^{2})}
\end{equation}
with $ \beta_{0} = 11 - 2n_{f}/3$, 
where the energy scale $Q^{2}$ is fixed to a given renormalization scale 
$\mu_{R}^{2}$.
The $\theta$ function "limits" the subtraction at the factorization scale 
$\mu_{F}^{2}$ since parton showers are limited by this scale.

We define the radiation factor as
\begin{equation}\label{eq:rad}
  f_{LL, i}(Q_{i}^{2},z)={\alpha_{s}(\mu_{R}^{2}) \over 2\pi}{P_{i}(z) \over z}
  {16 \pi^{2} \over Q_{i}^{2}},
\end{equation}
where $P_{i}(z)$ represents the leading-order Altarelli-Parisi splitting 
functions, given by
\begin{equation}\label{eq:split1}
	 P_{q \rightarrow qg}(z) = C_{F}{1+z^{2} \over 1-z},
\end{equation}
\begin{equation}\label{eq:split2}
	 P_{g \rightarrow gg}(z) = N_{C}{\{1- z(1-z)\}^{2} \over z(1-z)},
\end{equation}
\begin{equation}\label{eq:split3}
	 P_{g \rightarrow q{\bar q}}(z) = T_{R}\{z^{2}+(1-z)^{2}\},
\end{equation}
for the parton branches $q \rightarrow qg$, $g \rightarrow gg$, 
and $g \rightarrow q{\bar q}$, respectively.
The parameters are given as $C_{F} = 4/3$, $N_{C} = 3$, and $T_{R} = n_{f}/2$ 
with $n_{f} = 5$.
The parameter $Q^{2}$ is the squared momentum transfer of the radiation, 
defined as $Q^{2} = -t$.

The sum in Eq.~(\ref{eq:sub}) is taken over all possible sources of radiation 
in the picture of parton showers.
We can consider a unique source, $g \rightarrow q{\bar q}$, 
for the final-state quark in the process $qg \rightarrow A+q$, 
whereas there are two possible sources, $q \rightarrow qg$ and 
${\bar q} \rightarrow {\bar q}g$, 
that may produce the gluon in $q{\bar q} \rightarrow A+g$.
The $Q^{2}$ of the branch is different for the two possible branches 
in the latter case.
The orientation of the subsystem $A$ in its cm frame, 
for which the non-radiative ME is evaluated, 
may also be different since the boost and rotation due to the branch 
may be different.
There are two solutions of $Q^{2}$ for a give $p_{T}$ of the radiation:
\begin{equation}
  Q_{\pm}^{2} = \left\{ {1-z \over 2} \pm 
  \sqrt{ \left( {1-z \over 2} \right)^{2} - {p_{T}^{2} \over {\hat s}_{A+1}} } 
  \right\} {\hat s}_{A+1}.
\end{equation}
These two solutions correspond to the two possible branches that we consider, 
and the sum of the inverses of the two solutions is 
\begin{equation}
  {1 \over Q_{+}^{2}} + {1 \over Q_{-}^{2}} = {1-z \over p_{T}^{2} }.
\end{equation}
Therefore, since the $A$ + 0-jet MEs for the two possible branches become 
identical at the collinear limit, $p_{T}^{2}/{\hat s}_{A+1} \rightarrow 0$, 
the definition in Eq.~(\ref{eq:sub}) with Eq.~(\ref{eq:rad}) agrees 
with the definition in the previous paper~\cite{Kurihara:2002ne} 
at the collinear limit in this case.

The LLL subtraction works well for all processes supported in GR@PPA 2.8.
The divergences in the $A$ + 1-jet MEs are properly subtracted, 
and the remaining MEs are all finite.
Though we apply a small $p_{T}$ cut, $p_{T} >$ 1 GeV/$c$, 
to the additional parton in the $A$ + 1-jet processes for numerical stability, 
the cut effects are negligible because the differential cross sections 
converge to zero as $p_{T} \rightarrow 0$ in all the processes.

In order to achieve appropriate matching, 
we have to be careful about high $p_{T}$ behavior, too.
Since the radiation effect is separated by the energy scale $\mu_{F}$, 
the scale has to be consistently defined in the 0-jet and 1-jet processes.
Namely, we have to assign the same $\mu_{F}$ value to an $A$ + 1-jet event 
produced by the $A$ + 1-jet ME 
and to an $A$ + 0-jet event having the same topology 
after the application of a parton shower.
The problem is not serious in the case of single weak boson production 
processes.
We can choose a fixed $\mu_{F}$, typically, equal to the weak boson mass.
On the other hand, 
there may be many possible choices for diboson production processes.
For the 0-jet processes, the scale is frequently defined as
\begin{equation}\label{eq:muf1}
  \mu_{F}^{2} = {\bar m}_{V}^{2} + p_{T}^{2}, 
\end{equation}
where ${\bar m}_{V}^{2}$ is the average of the squared masses of the 
produced weak bosons, 
and $p_{T}$ is the transverse momentum of the hard interaction, 
$q{\bar q}' \rightarrow VV'$.
The definition in Eq.~(\ref{eq:muf1}) cannot be directly applied to 
1-jet processes since the diboson system has a transverse momentum, 
$q_{T}$, due to the radiation of the jet.
If we assume that the jet is radiated from one of the initial-state 
partons, 
we can define the 0-jet subsystem and define $\mu_{F}$ 
using Eq.~(\ref{eq:muf1}) in its cm frame.
However, this assumption is ambiguous and may be incorrect 
especially for those events having high $p_{T}$ radiation, 
for which the consistency has to be most seriously considered. 
There should not be a "correct" answer to this question.
Therefore, we adopt a simple definition in the sample program 
as one of the possible options, that is
\begin{equation}\label{eq:muf2}
  \mu_{F}^{2} = {\bar m}_{V}^{2} + \left| { \vec{p}_{T,1} - \vec{p}_{T,2} 
  \over 2 } \right|^{2}, 
\end{equation}
where $\vec{p}_{T,i}$ represents the transverse momentum vector 
of a weak boson.
Equation~(\ref{eq:muf2}) can be calculated only from the properties 
of weak bosons, and is in agreement with Eq.~(\ref{eq:muf1}) 
at the limit of $q_{T} \rightarrow 0$.

\section{Parton showers}
\label{sec:ps}

We provide three parton shower (PS) routines in GR@PPA 2.8: 
a forward-evolution initial-state PS (QCDPS), 
a backward-evolution initial-state PS (QCDPSb), 
and a final-state PS (QCDPSf).
All of them are based on the Sudakov form factor at the leading order, 
which is expressed as
\begin{equation}\label{eq:sudakov}
  S(Q_{1}^{2}, Q_{2}^{2}) = \exp\left[ - \int_{Q_{1}^{2}}^{Q_{2}^{2}}
  {dQ^{2} \over Q^{2}} \int_{\epsilon}^{1-\epsilon} dz \  
  {\alpha_{s}(Q^{2}) \over 2\pi}\ \sum_{i}P_{i}(z) \right] ,
\end{equation}
where $P_{i}(z)$ are the leading-order splitting functions 
in Eqs.~(\ref{eq:split1}-\ref{eq:split3}), 
and they are summed over all possible branches. 
The parameter $\epsilon$ cuts off the divergences in the splitting functions. 
We set the cutoff to be very small so that physical quantities should not be 
affected by the choice of this arbitrary parameter; 
$\epsilon=10^{-6}$ for the initial state and $10^{-3}$ for the final state 
as the default. 
A very small cutoff is applied to the initial state in order to make 
the total cross section stable at the level of one percent, 
though kinematical distributions of weak bosons are already stable 
at $\epsilon=10^{-3}$.

The PS branches are therefore ordered in $Q^{2}$.
For instance, in QCDPS, when we have a branch at $Q^{2} = Q_{1}^{2}$, 
the $Q^{2}$ of the next branch $Q_{2}^{2}$ is determined by solving 
the equation $S(Q_{1}^{2}, Q_{2}^{2}) = \eta$, 
using a random number $\eta$ that is uniformly generated in the range of 0 to 1.
$Q_{1}^{2}$ is derived from $Q_{2}^{2}$ in the same way in QCDPSf.
The splitting parameter $z$ is randomly determined in proportion to the 
relevant splitting functions $P_{i}(z)$.
We set the lower limit of $Q^{2}$ to be $Q_{0}^{2} =$ (4.6 GeV$)^{2}$, 
and the upper limit at the factorization scale $\mu_{F}^{2}$ 
if there is no other limitation\footnote{The lower limit is increased to 5.0 GeV 
since the GR@PPA 2.8.1 update.}. 

\subsection{Initial state}

The initial-state PS plays a crucial role in the matching method.
We use a forward-evolution PS, named QCDPS, as the primary tool for this purpose.
The forward evolution in the initial state is in general very inefficient, 
especially for those interactions which require tight constraints 
on the parton momenta after the evolution, 
such as narrow resonance productions,
because the final momentum is usually unpredictable in the forward evolution.
This problem has been solved by introducing the "$x$-deterministic" 
forward-evolution technique~\cite{Kurihara:2002ne}.
As discussed previously, parton showers can in principle replace the 
perturbative part of PDFs.
QCDPS realizes this concept in practical event generators.
When we use QCDPS, a PDF is employed only for setting the initial condition 
at the lower limit, $Q^{2} = Q_{0}^{2} $.

There is no ambiguity in the procedure for determining PS branches 
characterized by two parameters, $Q^{2}$ and $z$, 
on the basis of Eq.~(\ref{eq:sudakov}).
However, since the parameters are defined in the infinite-momentum frame, 
the correspondence of the determined parameters to the kinematical variables 
in a finite-momentum frame is not trivial.
We need to introduce a certain model to construct a practical PS.
In a previous paper~\cite{Odaka:2009qf}, 
we have pointed out that the definition 
of the $p_{T}$ of each PS branch is important in such models, 
and shown that the parton shower we have developed reproduces experimental 
data with surprisingly high precision. 
We apply the same kinematics model in GR@PPA 2.8, 
where the $p_{T}$ of each branch is "prefixed" according to the definition 
\begin{equation}\label{eq:pt}
  p_{T}^{2} = (1-z)Q^{2}.
\end{equation}
Since the branch kinematics do not affect the production cross section, 
this branch model is applied to events that are already unweighted, 
{\it i.e.}, they already have an event weight of $\pm 1$.
Though there is an option to include the kinematical smearing by PS 
in the cross section integration, this option remains experimental 
and it is not recommended for use in the present version.

Although the "$x$-deterministic" forward evolution technique significantly 
improves the generation efficiency, 
the application of QCDPS requires a long CPU time, 
compared to widely used backward-evolution parton showers.
Besides, the use of QCDPS requires another constraint. 
Since the leading-order (LO) Sudakov form factor is used, 
QCDPS cannot reproduce the evolution in next-to-leading order (NLO) 
or next-to-next-to-leading order (NNLO) PDFs, 
nor in the recently proposed modified leading-order (LO*) 
PDFs \cite{Sherstnev:2007nd,Lai:2009ne}.
If we use one of such PDFs for the initial condition at $Q^{2} = Q_{0}^{2}$, 
we will obtain an incorrect result on the parton distribution 
at larger energy scales relevant to the interactions of interest.
In order to overcome these problems in QCDPS, 
we also provide a backward-evolution PS named QCDPSb in GR@PPA 2.8.
The algorithm for generating PS branches is the same as PYSHOW in 
PYTHIA \cite{Bengtsson:1986gz}, the so-called "old model" of the PYTHIA PS. 
Though PYSHOW is known to give an unsatisfactory softer $p_{T}$ spectrum 
in weak boson productions,
we can expect QCDPSb to have a performance similar to that of QCDPS 
because we apply the same kinematics model.
However, QCDPSb does not strictly reproduce the evolution in PDFs, 
even when we compare it with LO PDFs.
We suggest that users should consider QCDPSb as a model 
for simulating initial-state hadronic activities, 
and check whether the desired properties are effectively reproduced by QCDPSb 
by comparing the results with those obtained using QCDPS.

\subsection{Final state}

The final-state PS in GR@PPA 2.8, named QCDPSf, remains experimental.
It is implemented mainly for consistency 
because the LHA user-process interface~\cite{Boos:2001cv}, 
which is used to pass the event information to other programs, 
has only one energy scale in each event.
When the QCD evolution is simulated down to $Q_{0}$ 
in the initial state by QCDPS or QCDPSb, 
the evolution has to be simulated to the same level also in the final state.
We describe the implemented final-state PS in some detail in this section 
because it is based on a new concept that we have learned in the development 
of the initial-state PS.

We have learned that consequences from discussions in the infinite-momentum 
frame, which may affect observable quantities, must be followed 
as strictly as possible in parton showers in a finite-momentum frame.
The effective definition of $p_{T}$ in each PS branch is most important  
because it determines the visible $p_{T}$ of the hard interaction system.
Another related consequence is 
that the evolution should not depend on the ordering of the branches.
However, these consequences are not consistent 
with energy-momentum conservation.
PS branches produce non-zero virtuality in at least one of the partons 
participating in each branch.
If the exact energy-momentum conservation is required, 
this virtuality necessarily affects the kinematics of preceding 
or subsequent branches; 
thus, the evolution becomes dependent on the ordering of branches 
and the originally applied $p_{T}$ definition is altered.
The "$p_{T}$-prefixed" kinematics model in QCDPS is designed to minimize 
the effect of this difficulty. 
However, there may be a kinematics model that can more strictly preserve 
the consequences in the infinite-momentum frame.

We have to violate energy-momentum conservation 
in order to realize the independence of PS branches.
If we disregard energy conservation, 
we can consistently determine the momenta of partons participating in a branch 
on the basis of arguments in the infinite-momentum frame.
Energy conservation can be restored by adjusting the overall scale 
of the momenta after completing the momentum determination for all branches.
This adjustment alters the momenta of the color-singlet products, 
such as weak bosons, as well.
However, we expect the alternation to be small and to have an insignificant 
effect on observable quantities.
We have developed QCDPSf on the basis of this concept.

\begin{figure}
\begin{center}
\includegraphics[scale=0.8]{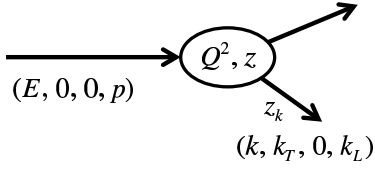}
\caption{\label{fig:branch}
An image of a PS branch in the final state.}
\end{center}
\end{figure}

We consider a final-state (time-like) PS branch of a parton 
having momentum $p$.
The branch is characterized by two parameters, $Q^{2}$ and $z$, 
as shown in Fig.~\ref{fig:branch}.
The two branch products are assumed to be massless.
The product that we are now going to study has momentum $k$, 
with a longitudinal component $k_{L}$ and a transverse component $k_{T}$ 
with respect to the direction of the mother;
{\it i.e.}, $k^{2} = k_{L}^{2} + k_{T}^{2} $. 
This branch product is assumed to have a "momentum fraction" $z_{k}$, 
which is either $z_{k}=z$ or $z_{k}=1-z$.

We define the momentum fraction in an infinite-momentum frame.
We consider a boost having a large Lorentz factor of $\tilde{\beta} \approx 1$ 
along the direction of the mother parton.
In order to carry out the Lorentz transformation consistently, 
we assume that the mother has an invariant mass $Q$, 
{\it i.e.}, it has the energy expressed as 
\begin{equation}\label{eq:emoth}
  E^{2} = p^{2} + Q^{2}.
\end{equation}
The kinematical variables in the infinite-momentum frame are denoted 
by tildes, 
{\it i.e.}, $\tilde{E}=\tilde{\gamma}(E+\tilde{\beta}p)$ and 
$\tilde{p}=\tilde{\gamma}(p+\tilde{\beta}E)$ 
with $\tilde{\gamma}^{2}=1/(1-\tilde{\beta}^{2})$.
The transverse momentum of the branch product is given by
\begin{equation}\label{eq:kt}
  k_{T}^{2} = z_{k}( 1 - z_{k} )Q^{2} = z( 1 - z )Q^{2}.
\end{equation}
This is invariant against the Lorentz boost.
On the other hand, using quantities in the infinite-momentum frame, 
the longitudinal momentum of the branch product can be expressed as
\begin{equation}\label{eq:kl1}
  k_{L}=\tilde{\gamma}(\tilde{k}_{L}-\tilde{\beta}\tilde{k})
  =\tilde{\gamma}\tilde{k}_{L}\left( 1-\tilde{\beta}
  \sqrt{ 1+{ k_{T}^{2} \over \tilde{k}_{L}^{2} } } \right).
\end{equation}
Since $k_{T}^{2}/\tilde{k}_{L}^{2} \ll 1$, 
it must be sufficient to take a first-order expansion of the square root as
\begin{equation}\label{eq:kl2}
  k_{L}=\tilde{\gamma}\tilde{k}_{L}
  \left\{ 1-\tilde{\beta}\left( 1+{k_{T}^{2} \over 2\tilde{k}_{L}^{2}} \right) 
  \right\}.
\end{equation}
Here, we define the momentum fraction $z_{k}$ as 
\begin{equation}\label{eq:zk}
  \tilde{k}_{L} = z_{k}\tilde{p}
\end{equation}
in the infinite-momentum frame. 
Since $\tilde{p}=\tilde{\gamma}(p+\tilde{\beta}E)$, 
$\tilde{k}_{L}$ can be expressed as 
$\tilde{k}_{L}=z_{k}\tilde{\gamma}(p+\tilde{\beta}E)$.
Substituting this relation in Eq.~(\ref{eq:kl2}) and taking $\tilde{\beta}=1$, 
we obtain the definition of the longitudinal momentum as 
\begin{equation}\label{eq:kl3}
  k_{L}={z_{k}(p+E) \over 2} - {k_{T}^{2} \over 2z_{k}(p+E)}.
\end{equation}
This definition has an undesirable feature that it may produce backward 
branches ($k_{L}<0$) 
which may form a jet-like structure in the backward direction.

The presence of these backward branches has already been discussed 
by Kato and Munehisa~\cite{Kato:1988ii} 
for the $e^{+}e^{-} \rightarrow q{\bar q}$ process. 
In principle, the QCD evolution for the final-state $q{\bar q}$ system 
can be completed only by applying a PS to either $q$ or ${\bar q}$ 
in an infinite-momentum frame ("single cascade").
Soft but not very collinear radiations are boosted backwards in the cm frame, 
and they reproduce radiations to be generated by the other quark.
Though theoretically the radiations become symmetric in the cm frame, 
the actual implementation of PS may produce a substantial asymmetry.
In order to overcome this technical difficulty, 
Kato and Munehisa proposed a "double cascade" scheme in which, 
from a study of the $e^{+}e^{-} \rightarrow q{\bar q}g$ differential 
cross section, 
they derived a condition to consistently separate the radiations. 
The condition gives a constraint on $z$ in each PS branch 
and effectively selects forward-going branches in the cm frame.
An independent application of a PS to $q$ and ${\bar q}$ with this condition 
is theoretically equivalent to the "single cascade" 
in the infinite-momentum frame, 
and technically it produces symmetric radiations in the cm frame. 

From the discussions on the "double cascade" scheme, 
we learn that the backward branches in Eq.~(\ref{eq:kl3}) can be interpreted 
as branches associated with a system compensating the color flow 
of the parton of interest, 
and that it is reasonable to reject these backward branches 
in an actual implementation of parton showers.
Hence, the question is: what is the compensating system in our case?
We cannot select any other parton or parton system in the PS 
since the branches must be independent of the ordering.
It would be natural to consider the rest of the whole colliding system 
as the compensating system 
because the color flow may be connected even with the remnants 
of beam collision.
By the way, we need not be very sensitive to the detailed definition 
of the "forward" and "backward" branches 
because most of the branches are concentrated in very forward and backward 
regions.

In the present version of QCDPSf, 
we apply the final-state PS in the cm frame of the hard interaction.
We take the momentum in this frame as $p$ in Eq.~(\ref{eq:kl3}), 
and retain only those branches having $k_{L} > 0$.
This selection yields the constraint 
\begin{equation}\label{eq:fwd1}
  z_{k}^{2}(p+E)^{2} > k_{T}^{2}. 
\end{equation}
Since $k_{T}$ is defined in Eq.~(\ref{eq:kt}), Eq.~(\ref{eq:fwd1}) can be
written as
\begin{equation}\label{eq:fwd2}
  z_{k} > {Q^{2} \over (p+E)^{2}+Q^{2}}. 
\end{equation}
The condition in Eq.~(\ref{eq:fwd2}) has to be satisfied by the two branch 
products.
Therefore, it is effectively a constraint on the lower-momentum product 
having $z_{k}<1/2$.
The right-hand side of Eq.~(\ref{eq:fwd2}) is always smaller than 1/2.
Thus, Eq.~(\ref{eq:fwd2}) does not give any clear constraint on $Q^{2}$.
However, it strongly limits the branch of low momentum partons.
The condition in Eq.~(\ref{eq:fwd2}) gives a constraint of $z_{k}>1/2$ 
at the limit $p/Q \rightarrow 0$; 
thus, no branch is allowed at this limit.
The constraint is rapidly relaxed as $p/Q$ increases.
The right-hand side of Eq.~(\ref{eq:fwd2}) is smaller than $10^{-2}$ 
for $p/Q>3$.

In the actual implementation, 
the condition in Eq.~(\ref{eq:fwd2}) is required each time 
a set of $Q^{2}$ and $z$ is generated, 
using the momentum $k = \sqrt{k_{L}^2+k_{T}^2}$ tentatively evaluated 
in the previous branch.
If the condition is not satisfied, 
the branch is discarded and the evolution proceeds 
by taking the discarded $Q^{2}$ as the maximum value of the next branch. 
The evolution starts from the factorization scale $\mu_{F}$, in principle.
If the $p_{T}$ of the parton with respect to the initial-state direction 
is smaller than $\mu_{F}$, 
the $p_{T}$ is chosen as the initial value. 
The evolution is terminated when the newly determined $Q^{2}$ becomes 
smaller than $Q_{0}^{2}$.
The full kinematics of the branches are determined after completing 
the generation of branches.
The branch direction is randomly chosen in azimuth in each branch 
in order to determine the three-momenta of the products.
The momenta of the final-state particles are then adjusted 
in order to restore the energy conservation. 
We adjust (decrease) the overall scale of the momenta of all particles 
that constitute the final state of the hard interaction 
so that the total invariant mass should match the value 
before the parton shower is applied. 
Weak bosons are treated as particles in order to preserve 
their invariant masses, 
and the momentum adjustment is carried out such that the decay directions 
in the cm frame of weak bosons remain unchanged.

The final-state PS can also be applied to the radiations produced 
by the initial-state PS.
If it is turned on, QCDPSf is applied when the kinematics 
of each initial-state branch is determined, 
with the maximum $Q^{2}$ set to the $p_{T}^{2}$ of the branch.
We assume that the radiation has a momentum equal to $p_{T}$.
The application of QCDPSf produces a non-zero mass of the radiation 
if it results in additional branches.
If any branch is added, 
the kinematics of the initial-state PS branch is adjusted 
by taking the produced mass into consideration, 
keeping the $p_{T}$ definition unchanged as far as possible. 
The momenta of branch products are then rotated and boosted 
so that the total momentum matches the adjusted radiation momentum.

\section{How to use}
\label{sec:run}

\subsection{Distribution package}

The program package is distributed as a gzipped tar file 
named {\tt GR@PPA-2.8.tgz}, 
which can be obtained from the GR@PPA Web page\footnote{\tt
http://atlas.kek.jp/physics/nlo-wg/grappa.html\#GRAPPA2.8}.
The compressed file can be expanded, for instance, by typing
\renewcommand{\baselinestretch}{0.78}
\begin{verbatim}
    tar zxf GR@PPA-2.8.tgz
\end{verbatim}
on UNIX systems.
When the file is expanded, 
users have a directory named {\tt GR@PPA-2.8} containing 
the following files and directories:

\begin{tabbing}
(miscellaneous files)\\
{\tt README} $\quad \quad \quad$ \= : \= readme file describing how to use 
this package,\\
{\tt VERSION-2.8} \> : \> file to show the version number,\\
\\
(files for setup)\\
{\tt config.input} \> : \> file to specify the configuration 
for the setup,\\
{\tt Install} \> : \> shell script for the installation,\\
{\tt config} \> : \> shell script to configure the setup,\\
{\tt config.perl} \> : \> Perl script called by {\tt config},\\
{\tt proc.list} \> : \> process list, which is referred to 
in {\tt config.perl},\\
\\
(GR@PPA framework)\\
{\tt grckinem} \> : \> source files of the framework,\\
{\tt basesv5.1} \> : \> BASES 5.1 source package with some customization,\\
{\tt chanel} \> : \> CHANEL source package to define the interaction model,\\
{\tt inc}  \> : \> directory containing common include files,\\
{\tt example} \> : \> directory to be used for the setup of sample programs,\\
{\tt diagrams} \> : \> directory containing PS files illustrating typical 
Feynman diagrams,\\
{\tt lib} \> : \> directory to store object libraries; initially empty,\\
\\
(process directories)\\
{\tt wjets} \> : \> $W$ production processes,\\
{\tt zjets} \> : \> $Z$ production processes,\\
{\tt diboson} \> : \> diboson ($W^{+}W^{-}$, $WZ$, $ZZ$) production processes.
\end{tabbing}

\subsection{Installation}

The installation of GR@PPA is easy on Unix/Linux systems 
on which the {\tt bash} shell and the Perl interpreter are available.
Users have to specify the user-dependent configuration for the installation 
in the file {\tt config.input}.
The items to be specified are 
the choice of a PDF library, compile and archive commands and their 
options, and the paths to external libraries and include directories.
Concerning the PDF library, it is most convenient to choose the built-in 
CTEQ6 PDF \cite{Pumplin:2002vw} 
for users who just want to test the included event generators, 
while it must be better to choose 
LHAPDF\footnote{{\tt http://projects.hepforge.org/lhapdf/}} 
\cite{Whalley:2005nh} 
if they want to do intensive studies.
Users can also choose the classical PDFLIB 
library\footnote{{\tt http://wwwasd.web.cern.ch/wwwasd/cernlib/}} 
in CERNLIB, if they want.
Fortran and C compilers are necessary for building GR@PPA libraries, 
while a C++ compiler is required only if users want to use 
RBOOK\footnote{A simple Fortran interface to the ROOT histogramming 
coded by W. Verkerke (NIKHEF). 
The source code was copied from the MC@NLO 3.4 package.}  
in sample programs. 
The external libraries are referred to for building sample programs.
Though they can be given later by editing the {\tt Makefile} 
for sample programs, 
it is better to specify them in {\tt config.input} in order to avoid problems 
arising from inconsistency in the compile conditions.
Some directory paths are also referred to in the configuration of sample 
programs.

A PYTHIA 6.4 library\footnote{The source code can be obtained from 
{\tt http://projects.hepforge.org/pythia6/}.} is necessary 
if users want to perform the simulations down to the hadron level.
When LHAPDF is chosen as the PDF library, 
it is required to specify the paths to its {\tt bin}, {\tt lib}, and 
PDF data directories separately because their location depends on 
the installation of LHAPDF.
The CERNLIB libraries are necessary to use HBOOK for histogramming 
in sample programs, 
while a ROOT\footnote{{\tt http://root.cern.ch/}} \cite{Antcheva20092499} 
library is necessary to use RBOOK.
Sample programs require at least one of the two libraries.
Unnecessary items can be left blank in {\tt config.input}.

Here, users can delete any process directories which they are not 
interested in.
The installation is done only for the retained processes.
If the configuration is given properly in {\tt config.input}, 
users can complete the installation of the libraries 
by invoking a shell script as
\renewcommand{\baselinestretch}{0.78}
\begin{verbatim}
    ./Install
\end{verbatim}
at the top directory of GR@PPA.
The script {\tt Install} creates a {\tt Makefile} by executing the script, 
{\tt config}, 
and then compiles the source files to install object libraries 
in the {\tt lib} directory by invoking the {\tt make} command.
It may be better to invoke {\tt ./config} and {\tt make} separately, 
instead of {\tt ./Install}, when users encounter a problem.
The script {\tt config} invokes {\tt config.perl} to create the {\tt Makefile} 
and to make other miscellaneous setups according to the configuration 
described in {\tt config.input}.
Parameters defined in {\tt config.input} are exported as environmental 
variables to be referred to in a Perl script {\tt config.perl}.
This export does not affect the shell environment of users 
because {\tt config} is executed on a separate shell.
The process directories are searched in {\tt config.perl} to add necessary 
descriptions to the {\tt Makefile} and some source files in {\tt grckinem} 
by referring to the process list in {\tt proc.list}.
No description is added for those processes 
for which the corresponding process directories are not found.
The installation is confined in the {\tt GR@PPA-2.8} directory
and nothing is created outside of it.
Therefore, users can completely clean up the installation by deleting 
the {\tt GR@PPA-2.8} directory, 
though some clean-up utilities are provided in the {\tt Makefile} 
created by the {\tt Install} script.

\subsection{Sample programs}

Sample programs are provided separately for each process.
For instance, those for the $Z$ production processes are placed 
in the {\tt zjets/example} directory.
There are two examples for the matched generation of the $Z$ + 0-jet 
and the $Z$ + 1-jet processes: 
{\tt z1j\_matched\_hbook} and {\tt  z1j\_matched\_rbook}.
The former uses HBOOK for histogramming and the latter uses RBOOK.
The produced histograms can be viewed and processed by PAW in CERNLIB 
in the former and by ROOT in the latter.
Moving to one of these directories, 
one can build and execute the program by invoking the commands 
\renewcommand{\baselinestretch}{0.78}
\begin{verbatim}
    ./Config
    make
    ./run
\end{verbatim}
The sample programs produce an event file using the LhaExt 
utility\footnote{{\tt http://atlas.kek.jp/physics/nlo-wg/grappa.html\#LhaExt}}.
The event data in the produced file can be further processed 
by PYTHIA~\cite{Sjostrand:2006za}.
A sample program is provided under the {\tt pythia} directory in each sample.
The parameters in PYTHIA are left unchanged from the default in the sample 
programs, 
except for {\tt PARP(67)= 1.0} and {\tt PARP(71)= 1.0} explicitly written 
in the source program {\tt Pythia.f}.
This setting should be applied when one uses the default "old" PS in PYTHIA.
One can build and execute the sample program by invoking the above three 
commands again in the {\tt pythia} directory.

The sample programs are fully relocatable once they are configured 
by executing the {\tt Config} scripts.
Users can copy the sample directory to anywhere within the same file system.
The configuration has to be done also in the {\tt pythia} directory 
before copying, if one wants to apply PYTHIA to the generated events.

\begin{table}[t]
\caption{Parameters to control the matching method and parton showers, 
which can be set in {\tt UPINIT}. "D" denotes the default choice 
in {\tt GRCINIT}.}
\label{tab:matching}
\begin{center}
\begin{tabular}{l l} 
\hline
Parameter & Description \\ 
\hline
{\tt matching} & switch for the matching method. (D = 0)\\
 & = 0 : no matching method is applied. \\
 & = 1 : the LLL subtraction is applied to 1-jet processes. \\
 \\
 {\tt ishower} & switch for the initial-state PS. (D = 0)\\
 & = 0 : no PS is applied. \\
 & = 1 : QCDPS is applied. \\
 & = 2 : QCDPSb is applied. \\
 & = 3 : QCDPS with a final-state PS for the radiations. \\
 & = 4 : QCDPSb with a final-state PS for the radiations. \\
\\
{\tt ishwfin} & switch for the final-state PS. (D = 0)\\
 & = 0 : no PS is applied. \\
 & = 1 : QCDPSf is applied. \\
 \hline
\end{tabular}
\end{center}
\end{table}

Each sample program contains four source files:
{\tt grappa.f}, {\tt grcpar.F}, {\tt upevnt.f}, and {\tt upinit.F}.
The file {\tt grappa.f} contains the main program of each sample, 
in which an analysis example is provided together with the necessary 
steering procedure.
Users can modify the execution of GR@PPA by customizing the three files,  
{\tt grappa.f}, {\tt grcpar.F} and {\tt upinit.F}.
Users will not need to edit {\tt upevnt.f} because it contains 
only one subroutine, {\tt UPEVNT}, which simply calls the GR@PPA steering routine 
{\tt GRCPYGEN} with {\tt mode} = 0 in the event generation stage.
The {\tt usr} directory in each sample program directory is used to store 
common utility routines and component files to be used to compose 
some sample program routines.

The file {\tt upinit.F} contains the subroutine {\tt UPINIT}.
This routine is called once before starting the event generation. 
The optimization of the random number generation and the cross section 
integration are done by calling {\tt GRCPYGEN} with {\tt mode} = 1 
at the end of {\tt UPINIT}.
Parameters controlling the matching method and parton showers have to 
be set in this routine. 
The allowed choices are listed in Table~\ref{tab:matching}.
The recommended combination is {\tt matching} = 1, {\tt ishower} = 3 or 4, 
and {\tt ishwfin} = 1.

\begin{table}[t]
\caption{Other GR@PPA parameters which users can set in {\tt UPINIT}. 
"D" denotes the default choice.}
\label{tab:upinit}
\begin{center}
\begin{tabular}{l l} 
\hline
Parameter & Description \\ 
\hline
{\tt CBEAM} & '{\tt PP}' for $pp$ collisions and '{\tt PAP}' for $p\bar{p}$ 
collisions. (D = '{\tt PP}')\\
\\
{\tt GRCECM} & CM energy of the beam collision in GeV. (D = 14000.0) \\
\\
{\tt GPTCUT} & Minimum $p_{T}$ in GeV for jets, except for those from 
weak-boson decays. \\
& (D = 20.0) \\ 
\\
{\tt GRCPTCUT(8)} & Minimum $p_{T}$ in GeV for each final state particle.
(D = 0.0) \\
\\
{\tt GETACUT} & Largest pseudorapidity in the absolute value for jets, 
except for \\
& those from weak boson decays. (D = 3.0) \\ 
\\
{\tt GRCETACUT(8)} & Largest pseudorapidity cut for each final state particle.
(D = 10.0) \\
\\
{\tt GRCONCUT} & Minimum separation cut in $\Delta R$ 
(= $\sqrt{\Delta\phi^{2}+\Delta\eta^{2}}$) for jets except \\
& for those from weak boson decays. (D = 0.4) \\ 
\\
{\tt GRCRCONCUT(8)} & Minimum separation cut in $\Delta R$ 
(= $\sqrt{\Delta\phi^{2}+\Delta\eta^{2}}$) for each final \\
& state particle. (D = 0.0) \\
\\
{\tt IGWMOD(20)} & Switch for the decay modes of $W$: 1 to activate and 0 to 
deactivate. \\
\\
{\tt IGZMOD(16)} & Switch for the decay modes of $Z$: 1 to activate and 0 to 
deactivate. \\
\\
{\tt IWIDCOR} & Switch for the decay widths correction for $W$ and $Z$. 
(D = 1)\\
& = 0: no correction; {\it i.e.}, the lowest-order values are used. \\
& = 1: corrected to match the values given in {\tt SETMAS} 
and branching ratios \\
& in {\tt GRCWBR} and {\tt GRCZBR}.\\
\\
{\tt GRCWBR(20)} & Branching ratios of $W$; valid only when 
{\tt IWIDCOR} = 2. \\
\\
{\tt GRCZBR(16)} & Branching ratios of $Z$; valid only when 
{\tt IWIDCOR}=2. \\
\\
{\tt IGJFLV(7)} & Switch for the jet flavors to be produced; 
1 to activate and 0 to deactivate. \\
& These are irrelevant to those from weak-boson decays.\\
\\
{\tt IGRCGEF} & Switch for the $Z$-$\gamma$ interference effect 
in $Z$ production processes;\\
& turned on if set to 1 and ignored if 0. (D=1) \\
\\
{\tt GRCCKM(3,3)} & Squared CKM matrix. \\
\\
{\tt IGAUGE} & Choice of the scheme to determine the electroweak parameters. \\
& The $G_{\mu}$ scheme is used as the default. (D=1) \\ 
\hline
\end{tabular}
\end{center}
\end{table}

Other parameters which can be set in {\tt UPINIT} are listed 
in Table~\ref{tab:upinit}.
Users have to set basic parameters concerning the beam conditions and 
the processes to be generated.
GR@PPA supports multiple process generation, 
in which generated events are randomly mixed according to the production 
cross sections.
The processes to generate have to be specified using a parameter {\tt NPRUP} 
and an array {\tt LPRUP} defined in the LHA user-process interface \cite{Boos:2001cv}. 
In order to apply the matching method, 
we have to set {\tt NPRUP} = 2 and set the process numbers 
for the 0-jet and 1-jet processes in {\tt LPRUP(1)} and {\tt LPRUP(2)}.
Though the appropriate numbers are already set in the sample programs,
users can find the process numbers defined in GR@PPA in {\tt grcpar.F} 
and {\tt proc.list}.

The PDF to be used for evaluating the cross sections and/or for setting 
the initial condition of QCDPS also has to be specified in {\tt UPINIT}.
Users have to modify appropriate lines according to the choice of the library 
in {\tt config.input}.
Users do not need to care about those parts irrelevant to the choice 
because they are discarded in the preprocessing before compilation. 
The setting is simplified for LHAPDF
because a symbolic link, {\tt PDFsets}, pointing to the PDF data directory 
given in {\tt config.input} is created by {\tt Config} 
in the sample program directory when LHAPDF is chosen.
Refer to the manual of the chosen PDF library for more details.

In addition, users can choose decay modes of weak bosons 
and customize kinematical cuts in {\tt UPINIT}. 
However, it should be noted that the cuts are 
applied to the quantities before applying the parton showers, 
and cuts on the weak-boson decay products never dramatically improve 
the generation efficiency. 
Moreover, any change concerning the jets may affect the performance 
of the matching method.
Therefore, we recommend that those parameters relevant to the kinematical 
cuts should be kept unchanged from the preset values in {\tt UPINIT} 
of the sample programs.
The default setting corresponds to the no-cut condition.
If users want to apply any cuts relevant to a restricted detection condition, 
it would be better to apply them in the analysis part in {\tt grappa.f} 
and {\tt Pythia.f}.
Refer to the previous reports~\cite{Tsuno:2002ce,Tsuno:2006cu}
for more details of the parameters in Table~\ref{tab:upinit}.

\begin{table}[t]
\caption{Parameters which can be set in {\tt GRCPAR}. 
"D" denotes the default choice.}
\label{tab:grcpar}
\begin{center}
\begin{tabular}{l l}
\hline
Parameter & Description\\ 
\hline
{\tt ICOUP} & Choice of the renormalization scale. \\
 & = 1 : $\sqrt{\hat{s}}$ of the hard interaction. \\
 & = 2 : average of squared transverse mass ($<m_{T}^{2}>$). \\
 & = 3 : total squared transverse mass ($\sum m_{T}^{2}$). \\
 & = 4 : maximum squared transverse mass ($max$ $m_{T}^{2}$). \\
 & = 5 : fixed value. Set {\tt GRCQ} in GeV. \\
 & = 6 : user defined scale. Set {\tt GRCQ} in the subroutine 
{\tt GRCUSRSETQ}. \\
\\
{\tt IFACT} & Choice of the factorization scale. (D=0) \\
 & The definition is the same as {\tt ICOUP}. \\
 & If {\tt IFACT} = 0, the same value as the renormalization scale is used. \\
 & In case {\tt IFACT} = 5 or 6, set {\tt GRCFAQ} in GeV. \\
\\
{\tt GRCFILE} & Output file name for the BASES integration. \\
\\
{\tt IBSWRT} & Mode selection for BASES integration: 
= 0 for calling the BASES \\
& integration, and 1 for skipping. (D=0) \\ 
& If {\tt IBSWRT} = 1, the file defined in {\tt GRCFILE} is used. \\
\\
{\tt NCALL} & Number of sampling points in each step of the iterative grid \\
& optimization in BASES. \\
\\
{\tt mfbsps} & Multiplication factor to the number of sampling points 
in BASES, \\
& to be applied in the 2nd stage when QCDPS is used.\\
\\
{\tt INPFL} & Number of flavors used in the coupling calculation and PDF.
(D=5)\\ \hline
\end{tabular}
\end{center}
\end{table}

The file {\tt grcpar.F} contains routines which are frequently called 
during the execution of GR@PPA.
This file is created by {\tt config.perl} and copied to a sample program 
directory by {\tt Config} in each sample.
The subroutine {\tt GRCPAR} in this file defines parameters 
depending on the process.
The parameters that users are allowed to change are listed 
in Table~\ref{tab:grcpar}.
These parameters have to be given for every process.
For single weak-boson production processes, 
it must be best to choose fixed values for the renormalization scale 
and the factorization scale 
by setting {\tt ICOUP} = 5 and {\tt IFACT} = 5.
The parameters {\tt GRCQ} and {\tt GRCFAQ} have to be given explicitly 
in {\tt GRCPAR} in this case.
On the other hand, there may be many possible definitions for diboson 
production processes.
We have chosen {\tt ICOUP} = 6 and {\tt IFACT} = 6, and as an example 
we have defined the scales according to Eq.~(\ref{eq:muf2}) 
in the subroutine {\tt GRCUSRSETQ},
in which we take an identical definition for the two scales.

The parameter {\tt NCALL} defines the number of sampling points 
in each step of the BASES iteration. 
The numbers preset in the sample programs are optimized for the generation 
in the LHC condition.
Users need to optimize them for other conditions.
The guiding principle is to achieve a statistical accuracy of 0.2\% 
in the first stage of BASES 
and to eliminate any unexpected jump (increase) in accuracy in each iteration.

In order to implement QCDPS, 
a small modification has been applied to the BASES library included 
in the package. 
Because the random numbers in QCDPS are not handled by BASES,
the application of QCDPS appears as a random distribution of the event weight 
for each sampling point.
Therefore, the application of QCDPS degrades the statistical accuracy of BASES.
In order to compensate for this degradation, 
the number of sampling points that is originally determined by {\tt NCALL} 
is multiplied by a factor, {\tt mfbsps}, when QCDPS is applied.
In the first stage of BASES, the distribution of random numbers in BASES 
is optimized by adjusting a multi-dimensional grid for the sampling.
QCDPS is not applied there even if its use is required. 
Instead, the given PDF is directly referred to 
since QCDPS has nothing to do with this optimization.
QCDPS is applied and the multiplication to the number of sampling points 
is activated in the second stage, 
where the maximum event weight is determined in each hypercube 
by increasing the statistics.
Therefore, when the use of QCDPS is required, 
the second stage of BASES consumes longer CPU time than the first stage. 
Users can check the performance of QCDPS for the longitudinal QCD evolution 
by comparing the cross section results from the two stages of BASES.
A statistical accuracy of about 0.5\% would be enough in the second stage 
of BASES when QCDPS is applied.
The factor {\tt mfbsps} should be increased if an unnatural structure 
is observed in the rapidity distribution of the hard interaction system.

The subroutine {\tt SETMAS} in {\tt grcpar.F} defines basic properties 
of particles and interactions, 
such as masses, decay widths, and coupling constants.
Users may change these parameters if they want.
In addition, users can apply their own cuts to the events to be generated 
by customizing the subroutine {\tt GRCUSRCUT} in {\tt grcpar.F}.
They can define any cuts which cannot be accomplished by using the parameters 
in Table~\ref{tab:upinit}. 
However, as we have already mentioned, 
it is not recommended to apply any cut relevant to jets in this version 
because they may deteriorate the performance of the jet matching.
It is safe to apply those cuts relevant to Lorentz-invariant quantities 
of weak bosons.
For instance, it is necessary to define, at least, the lower limit 
of the $Z$-boson mass when the interference with the photon exchange 
is turned on ({\tt IGRCGEF} = 1).
The subroutine {\tt GRCUSRSETQ} defines the energy scales, 
{\tt GRCQ} and {\tt GRCFAQ}, when {\tt ICOUP} = 6 and/or {\tt IFACT} = 6. 
In {\tt GRCUSRCUT} and {\tt GRCUSRSETQ}, 
users can access the internal event information through the arrays, 
{\tt PGRC} and {\tt PLGRC}. 
Some utility functions are also available there.
Refer to the comments in {\tt grcpar.F} for details.
The internal particle numbering can be found in Feynman diagrams illustrated 
in the figures in the {\tt diagram} directory under the top directory.
Please ignore the subroutine {\tt grclabcut} because its implementation is 
still in an experimental phase.

The simulation can proceed to the hadron level by applying PYTHIA 
to the generated events. 
The sample programs that we provide in the {\tt pythia} directory 
employ the so-called "old" PS as the default. 
If we change it to the "new" PS by setting, for instance, {\tt MSTP(81) = 21}, 
PYTHIA issues many warnings and sometimes the execution hangs.
This would not be a problem caused by the combined use of PYTHIA and GR@PPA 
because similar warnings frequently appear even in stand-alone event 
generations. 
Though the reason is yet to be intensively investigated, 
the problem becomes less severe if we turn off parton showers in GR@PPA.
As we have shown in a previous report \cite{Odaka:2007gu}, 
the PYTHIA "new" PS shows transverse activities very similar to QCDPS.
If users are eager to use the "new" PS, 
it may be better to turn off all parton showers in GR@PPA 
by setting {\tt matching = 1}, {\tt ishower = 0}, and {\tt ishwfin = 0}, 
and pass the generated hard interaction events directly to PYTHIA.
In this case, the given factorization scale value ({\tt GRCFAQ}) is passed 
to PYTHIA as the energy scale in the LHA user-process interface.
When the "new" PS is applied, 
users have to comment out the line setting {\tt PARP(67)} 
in the main program {\tt Pythia.f}. 

Users may encounter another problem in the PYTHIA simulation 
when they turn on the parton showers in GR@PPA. 
The parton showers sometimes generate very large number of particles, 
and PYTHIA issues a warning and sometimes generate severe errors 
when the total number of particles passed through the LHA interface exceeds 80, 
even though up to 500 particles are allowed in the LHA interface.
Because it is very rare for the total number of particles to exceed 80 
even in the LHC condition with the full implementation of parton showers 
in GR@PPA, 
it may be better to skip such events in the subroutine {\tt UPEVNT} 
included in {\tt Pythia.f}.

The HERWIG PS~\cite{Corcella:2000bw,Corcella:2002jc} seems to be incompatible 
with the parton showers in the present version of GR@PPA. 
The execution is immediately terminated with an error.
Contrary to the case of the PYTHIA "new" PS, 
it would not be a good choice to fully replace the GR@PPA parton showers 
with the HERWIG PS 
because there is a marked difference between them \cite{Odaka:2009qf}.
Therefore, we do not officially support the combination with HERWIG 
in the present version.

\section{Simulation results}
\label{sec:res}

In this section, we show some results to verify the performance of GR@PPA.
The presented results are all obtained with the default setting in the sample 
programs, except for the number of events to generate.
Concerning the decay modes of the weak bosons, only the modes 
$W \rightarrow e\nu$ and $Z \rightarrow e^{+}e^{-}$ are activated.
The presented quantities are extracted after applying the PYTHIA simulation 
to the generated events.

\subsection{$W$ and $Z$ productions}

\begin{figure}
\begin{center}
\includegraphics[scale=0.8]{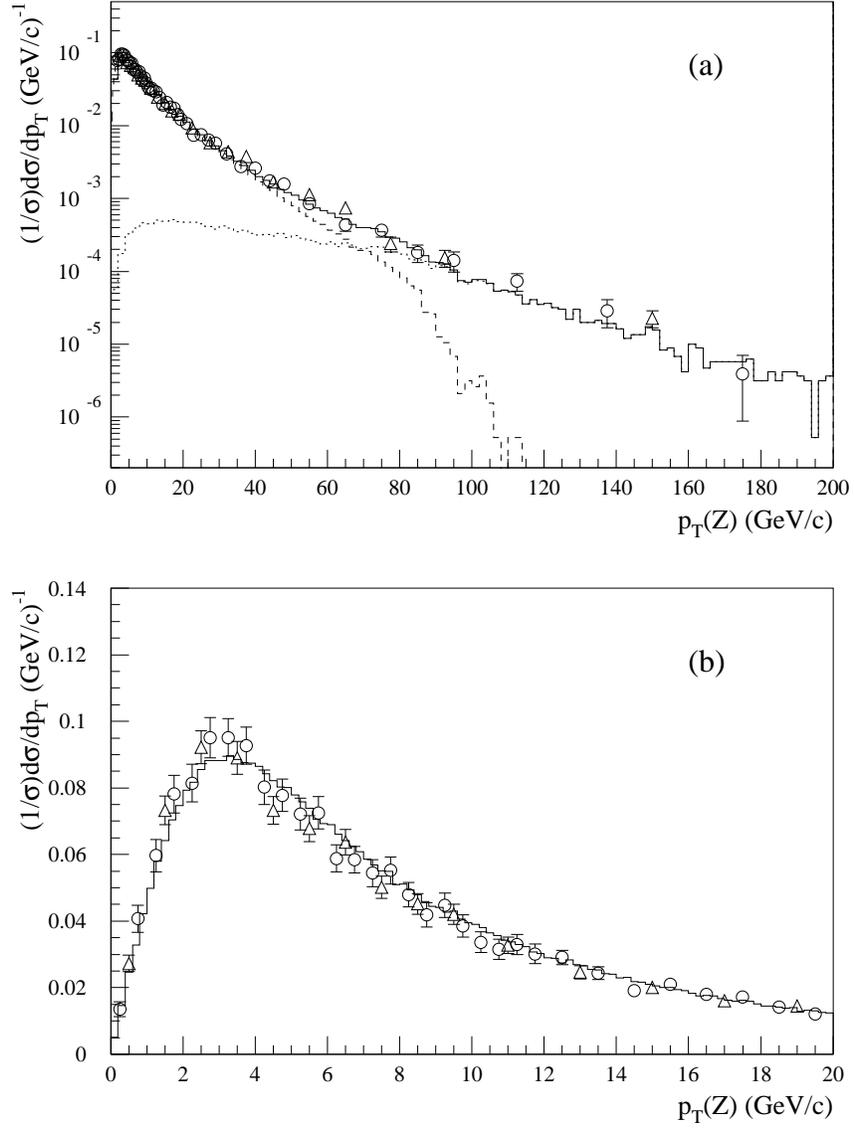}
\caption{\label{fig:zpt-tev}
$p_{T}$ spectrum of $Z$ bosons at Tevatron Run 1, 
$p\bar{p}$ collisions at a cm energy of 1.8 TeV.
The GR@PPA simulation (histograms) is compared with the measurements 
by CDF \cite{Affolder:1999jh} (circles) and D0 \cite{Abbott:1999wk} (triangles).
Together with a result covering the $p_{T}$ range up to 200 GeV/$c$ (a), 
a result to cover the range up to 20 GeV/$c$ (b) is presented 
to show the low-$p_{T}$ behavior.
In addition to the summed spectrum (solid), 
the spectra of events from the $Z$ + 0-jet (dashed) and $Z$ + 1-jet (dotted) 
processes are separately shown for the GR@PPA simulation in (a).}
\end{center}
\end{figure}

We have shown in a previous report~\cite{Odaka:2009qf} that the simulation 
employing our matching method and QCDPS reproduces the $p_{T}$ spectrum 
of the $Z$ bosons measured at Tevatron 
experiments~\cite{Affolder:1999jh,Abbott:1999wk,Abazov:2007nt} 
with a very good precision.
Figure~\ref{fig:zpt-tev} shows the result of a simulation on the same quantity, 
in which, together with QCDPS, QCDPSf is applied to partons from the hard 
interaction and those radiated in QCDPS, as described in the previous section.
The factorization scale ($\mu_{F}$) and the renormalization scale ($\mu_{R}$) 
are taken to be equal to the $Z$-boson mass (91.17 GeV/$c^{2}$),  
and the built-in CTEQ6L1 is used for PDF.
PYTHIA 6.421 is applied to the generated events in order to add simulations 
at lower energy scales, 
such as the primordial $k_{T}$ effect, hadronization, and decays.
Though this PYTHIA version is slightly newer than that used 
in the previous study, 
no significant change is observed in the quantities 
that we are currently interested in.
Together with the summed spectrum, 
the contributions from the $Z$ + 0-jet process and the LLL-subtracted 
$Z$ + 1-jet process are separately shown in the figure.
We can see that the two components are smoothly combined.
The $Z$ + 0-jet events are strongly suppressed at $p_{T} > \mu_{F}$ 
and the spectrum is totally determined by the $Z$ + 1-jet events 
at high $p_{T}$.
The LLL subtraction does not sharply suppress the $Z$ + 1-jet events 
at $p_{T} \lesssim \mu_{F}$ 
because non-logarithmic contributions are comparable around 
$Q^{2} = \mu_{F}^{2}$.
We can see that the $Z$ + 0-jet component overwhelms the remaining 
non-logarithmic $Z$ + 1-jet component at low $p_{T}$,
and the non-logarithmic component converges to zero as $p_{T} \rightarrow 0$.
Though the summed spectrum is a little bit softer compared to the previous 
result because of the application of QCDPSf, 
the simulation is still in good agreement with the measurements.

\begin{figure}
\begin{center}
\includegraphics[scale=0.8]{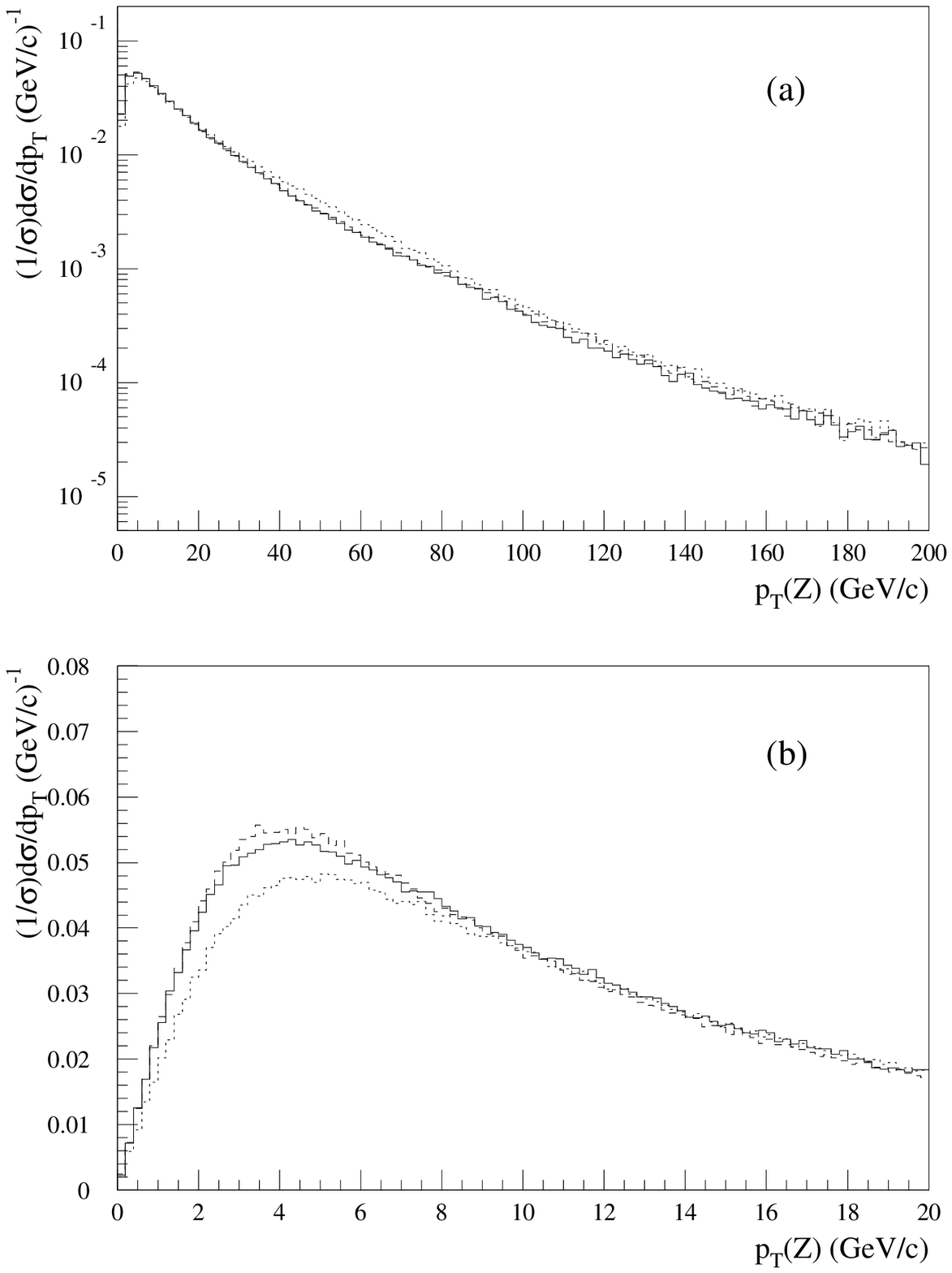}
\caption{\label{fig:zpt-lhc}
$p_{T}$ spectrum of $Z$ bosons at LHC, $pp$ collisions at a cm energy 
of 14 TeV.
The GR@PPA simulation (solid) is compared with the predictions 
from PYTHIA 6.421 (dashed) and HERWIG 6.510 (dotted).}
\end{center}
\end{figure}

Figure~\ref{fig:zpt-lhc} shows the $Z$-boson $p_{T}$ spectrum expected 
for the LHC condition, 
proton-proton collisions at a cm energy of 14 TeV, 
with the same choice of $\mu_{F}$, $\mu_{R}$ and PDF.
The GR@PPA simulation is compared with those from 
PYTHIA 6.421~\cite{Sjostrand:2006za} and 
HERWIG 6.510~\cite{Corcella:2000bw,Corcella:2002jc}.
The compared PYTHIA simulation employs its built-in event generator 
with the "new" PS model, 
and a primordial $k_{T}$ effect is added to the HERWIG simulation, 
as described in our previous report~\cite{Odaka:2009qf}.
The $Z$-boson invariant mass is required to be greater than 60 GeV/$c^{2}$ 
in all simulations.
The overall tendencies are the same as those observed for the simulations 
in the Tevatron condition~\cite{Odaka:2009qf}.
The GR@PPA simulation and the PYTHIA new-PS simulation are nearly identical 
in the whole $p_{T}$ range.

\begin{figure}
\begin{center}
\includegraphics[scale=0.65]{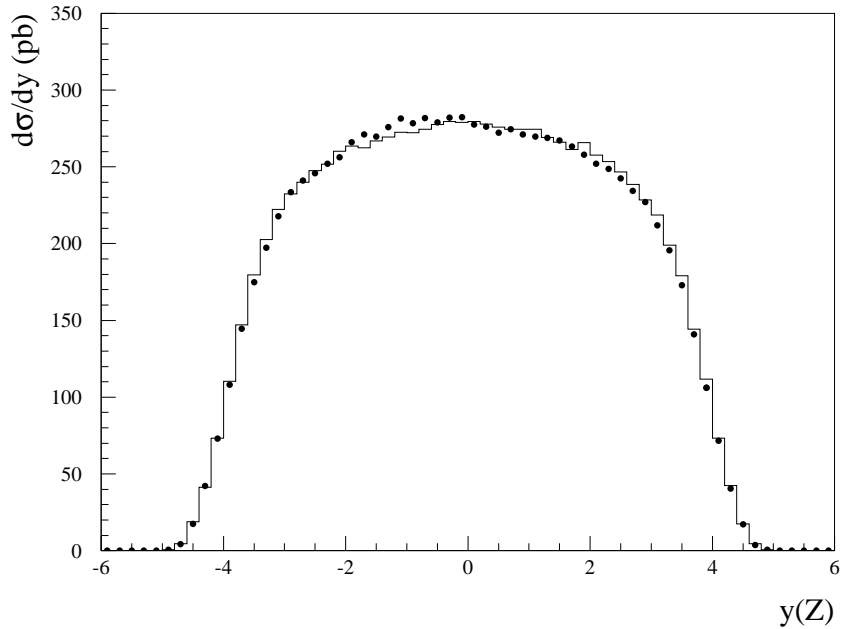}
\caption{\label{fig:zrap-lhc}
Rapidity distribution of $Z$ bosons at LHC.
The GR@PPA simulation employing QCDPS (plots) is compared with 
that employing QCDPSb (histogram). 
The QCD evolution is simulated using the parton shower in the former, 
while PDF is directly referred to in the latter.
Note that compared are the absolute values from the two simulations 
for $Z \rightarrow e^{+}e^{-}$ decays with $m_{e^{+}e^{-}} \geq 60$ 
GeV/$c^{2}$.}
\end{center}
\end{figure}

The rapidity distribution of $Z$ bosons is predominantly determined 
by the parton momentum distribution inside protons.
The distributions from the simulation using QCDPS and that using QCDPSb 
are compared in Fig.~\ref{fig:zrap-lhc}.
The longitudinal momentum evolution is simulated by the parton shower 
in the former, 
while the evolution in the PDF is directly referred to in the latter. 
The built-in CTEQ6L1 is used for the PDF, and the $Z$ bosons are tagged 
by $Z \rightarrow e^{+}e^{-}$ decays with an invariant mass 
constraint of $m_{e^{+}e^{-}} \geq 60$ GeV/$c^{2}$ in both simulations.
The two simulations are in good agreement.
Note that the absolute values of the differential cross section are 
compared in Fig.~\ref{fig:zrap-lhc}.
The agreement shows that the evolution by the parton shower in QCDPS 
satisfactorily reproduces the analytical evolution in CTEQ6L1.

\begin{figure}
\begin{center}
\includegraphics[scale=0.65]{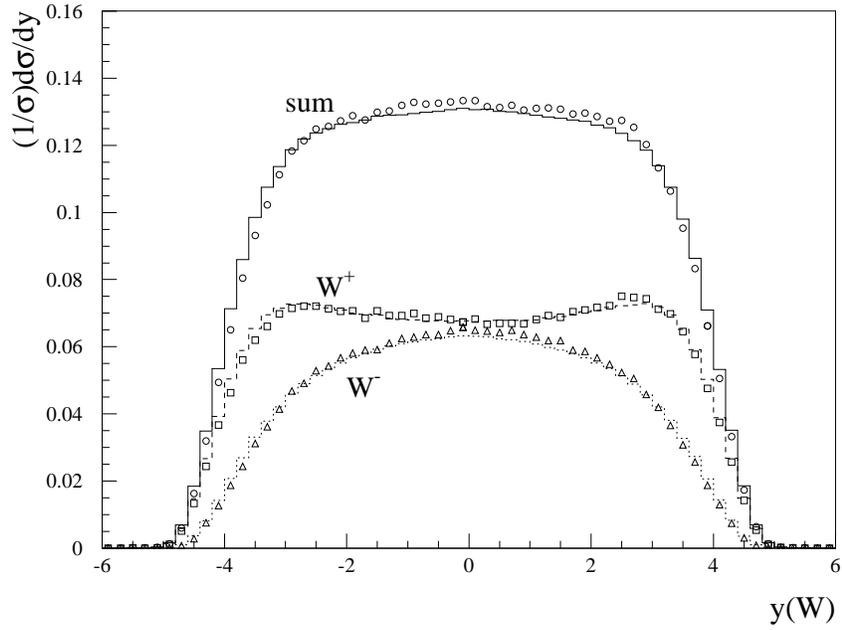}
\caption{\label{fig:wrap-lhc}
Rapidity distribution of $W$ bosons at LHC.
The GR@PPA simulation (plots) is compared with the PYTHIA simulation 
(histograms).
CTEQ6L1 is used for PDF in both simulations.}
\end{center}
\end{figure}

\begin{figure}
\begin{center}
\includegraphics[scale=0.65]{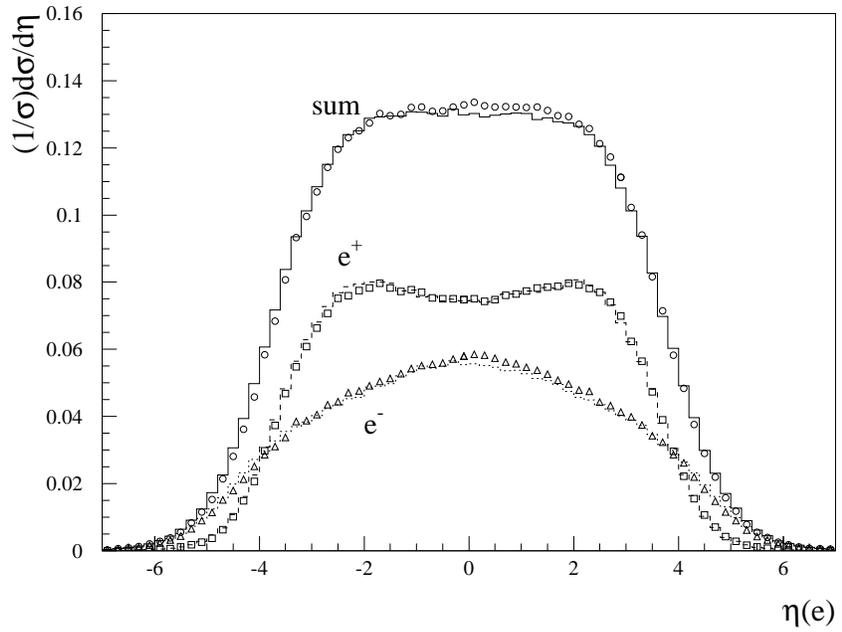}
\caption{\label{fig:weeta-lhc}
Pseudo-rapidity distribution of electrons from $W$-boson decays at LHC.
The GR@PPA simulation (plots) is compared with the PYTHIA simulation 
(histograms).
CTEQ6L1 is used for PDF in both simulations.}
\end{center}
\end{figure}

\begin{figure}
\begin{center}
\includegraphics[scale=0.65]{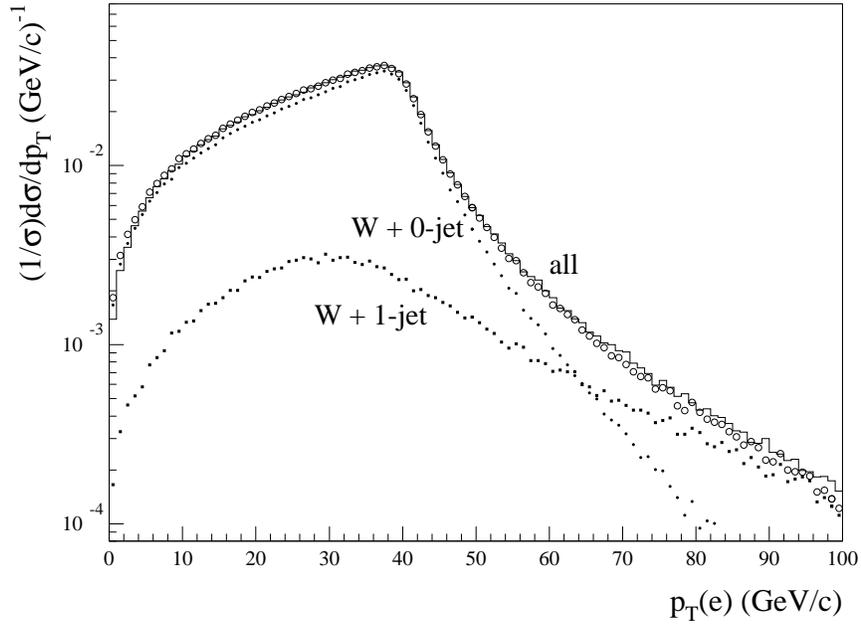}
\caption{\label{fig:wept-lhc}
$p_{T}$ distribution of electrons from $W$-boson decays at LHC.
The GR@PPA simulation (circles) is compared with the PYTHIA simulation 
employing the "new" PS model (histogram).
The contributions from the $W$ + 0-jet and the LLL-subtracted $W$ + 1-jet 
processes are separately shown as well for the GR@PPA simulation.}
\end{center}
\end{figure}

\begin{figure}
\begin{center}
\includegraphics[scale=0.65]{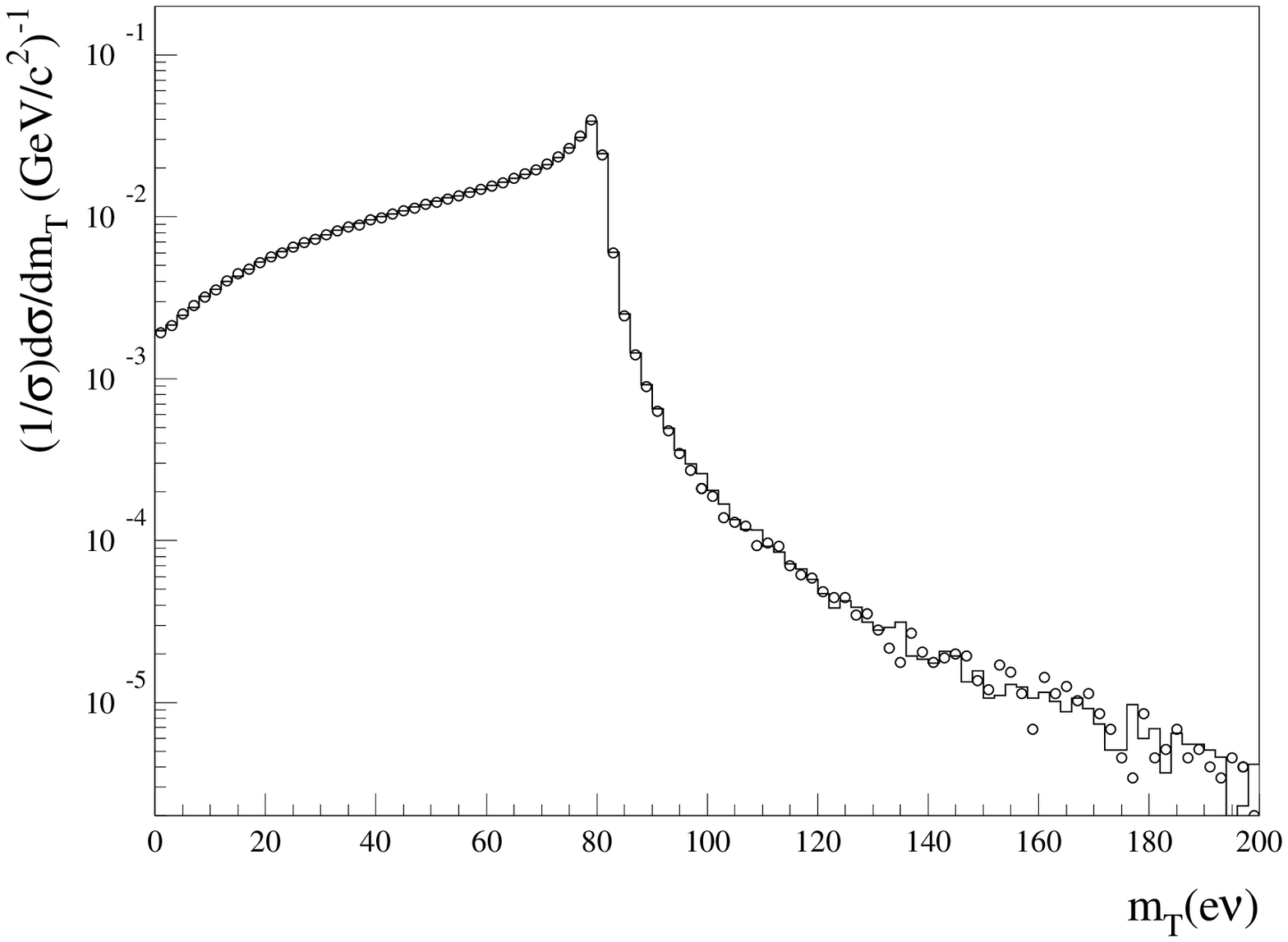}
\caption{\label{fig:wemt-lhc}
Transverse mass ($m_{T}$) distribution of $W$ bosons at LHC,
where $m_{T}$ is calculated from the momenta of the products from electronic 
decays, $W \rightarrow e\nu$.
The GR@PPA simulation (circles) is compared with the PYTHIA simulation 
(histogram).}
\end{center}
\end{figure}

Hereafter, we present some results concerning $W$ boson production 
in the LHC condition, 
where we set the renormalization and factorization scales to the $W$-boson 
mass (80.42 GeV), and we use the built-in CTEQ6L1 for the PDF.
The simulation results are compared with the predictions from PYTHIA.  
The PYTHIA simulation employs the "new" PS model 
and uses CTEQ6L1 in LHAPDF for the PDF with the help of 
LHAGLUE\footnote{See the LHAPDF online manual at 
{\tt http://projects.hepforge.org/lhapdf/manual\#tth\_sEc3.2}.}.
We expect that this PYTHIA simulation is at the same level as GR@PPA, 
except for the overall normalization. 
PYTHIA does not add the $W$ + 1-jet contribution to the total cross section.
Hence, the results are compared in the form of the relative shape of 
distributions.

The rapidity distribution is separately shown in Fig.~\ref{fig:wrap-lhc} 
for $W^{+}$ and $W^{-}$ together with the summed distribution.
Figure~\ref{fig:weeta-lhc} shows the pseudo-rapidity ($\eta$) distributions 
of the $e^{+}$ and $e^{-}$ from $W^{+}$ and $W^{-}$ decays, respectively.
The summed $p_{T}$ distribution is shown in Fig.~\ref{fig:wept-lhc}.
The contributions from the $W$ + 0-jet and $W$ + 1-jet processes are also
separately shown in Fig.~\ref{fig:wept-lhc}.
We can see that the tail of the distribution extends to a very high $p_{T}$ 
region owing to the existence of hard radiations in the $W$ + 1-jet process.
The distribution is slightly smeared due to photon radiations 
because we have taken final stable electrons in the PYTHIA event record.
Figure~\ref{fig:wemt-lhc} shows the transverse mass distribution. 
Here, we define the transverse mass ($m_{T}$) as
\begin{equation}\label{eq:mt}
  m_{T}^{2} = 2 \left( p_{T}^{(e)}p_{T}^{(\nu)} - \vec{p}_{T}^{(e)} 
\cdot \vec{p}_{T}^{(\nu)} \right), 
\end{equation}
where $\vec{p}_{T}^{(e)}$ and $\vec{p}_{T}^{(\nu)}$ denote the transverse 
momentum vector of the electron and neutrino, respectively, 
with $p_{T}^{(e)}$ and $p_{T}^{(\nu)}$ representing their absolute values. 
The tail to large $m_{T}$ values is an off-shell effect of $W$ bosons.
The measurement of this tail provides us with an opportunity to directly 
measure the $W$-boson decay width.

The results from GR@PPA and PYTHIA are in good agreements 
in all distributions shown in Figs.~\ref{fig:wrap-lhc}-\ref{fig:wemt-lhc}. 
The agreement is not trivial because there are many differences 
in the simulation technique, 
not only in the matching method but also in the treatment of $W$ bosons.
PYTHIA generates $W$ bosons based on the on-shell approximation.
The finite decay width and asymmetry in the decay angle are attached 
afterwards.
On the other hand, the off-shell effects and decay dynamics are 
included in the matrix elements in the GR@PPA simulation.
It should also be noted that GR@PPA employs only the $u\bar{d} \rightarrow 
W^{+}$ interaction as the base process. 
The $W^{-}$ production and other flavor contributions are derived 
by applying the charge conjugation and flavor exchanges to the base process.
The agreement that we can see in Figs.~\ref{fig:wrap-lhc}-\ref{fig:wemt-lhc} 
implies that these techniques are properly implemented in both simulations.

\subsection{Diboson productions}

\begin{table}
\caption{Benchmark cross sections in pb for $VV$ + $N$ jets processes, 
correcting Table~4 in the previous report~\cite{Tsuno:2006cu}.}
\label{tab:vvres}
\begin{center}
\begin{tabular}{c c c c}
\hline
\multicolumn{4}{l}{Tevatron Run II}\\ 
  $N$ jets & $W^{+}W^{-}$ & $WZ$ & $ZZ$ \\
\hline
0 & 7.06(1)$\times$10$^{-2}$ 
  & 6.10(2)$\times$10$^{-3}$ 
  & 7.71(1)$\times$10$^{-4}$ \\

1 & 1.769(3)$\times$10$^{-2}$ 
  & 1.608(4)$\times$10$^{-3}$ 
  & 1.860(3)$\times$10$^{-4}$ \\
\hline
\multicolumn{4}{l}{LHC}\\ 
  $N$ jets & $W^{+}W^{-}$ & $WZ$ & $ZZ$ \\
\hline
0 & 4.91(1)$\times$10$^{-1}$
  & 4.66(2)$\times$10$^{-2}$
  & 5.40(1)$\times$10$^{-3}$ \\

1 & 3.275(7)$\times$10$^{-1}$
  & 4.57(1)$\times$10$^{-2}$
  & 3.031(6)$\times$10$^{-3}$ \\ 
\hline
\end{tabular}
\end{center}
\end{table}

Some bugs have been identified in the diboson production processes 
included in the previous version of GR@PPA, 
and have been fixed for the $VV$ + 0-jet and $VV$ + 1-jet production 
processes included in the present version. 
As a result, some numbers presented in Table~4 
in the previous report~\cite{Tsuno:2006cu} were incorrect. 
The corrected numbers are presented in Table~\ref{tab:vvres}. 

\begin{figure}
\begin{center}
\includegraphics[scale=0.65]{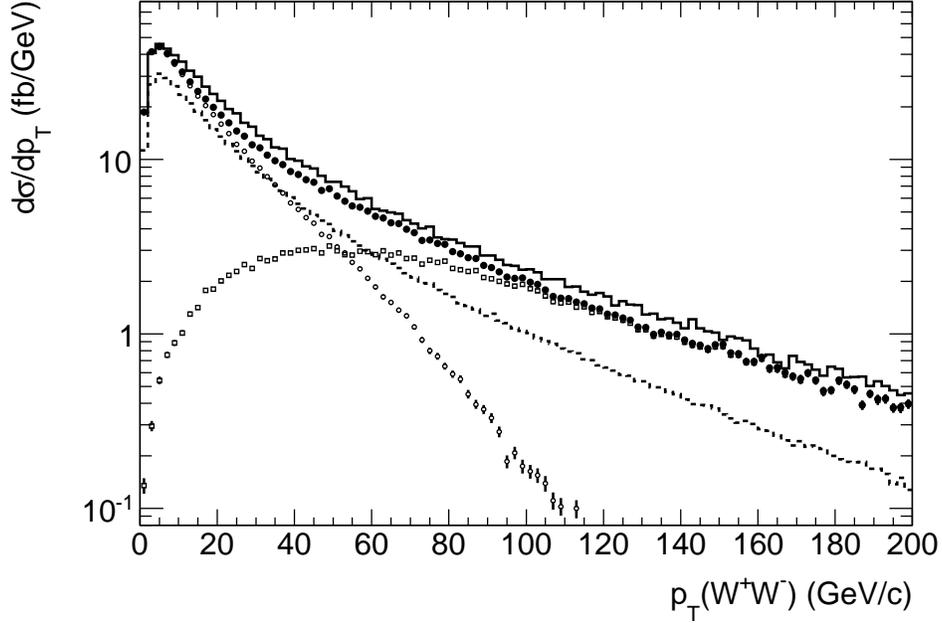}
\caption{\label{fig:ww_ptvv}
$p_{T}$ distribution of the diboson system in the $W^{+}W^{-}$ production 
process at LHC. 
The results are shown in the absolute values of the cross section.
The GR@PPA result is separately shown for the 0-jet (open circles) 
and LLL-subtracted 1-jet (open squares) contributions 
together with the summed result (filled circles).
The result is compared with the predictions from MC@NLO (solid histogram) 
and PYTHIA (dashed histogram).
}
\end{center}
\end{figure}

Hereafter, we show some results on diboson production processes.
Figure~\ref{fig:ww_ptvv} shows the $p_{T}$ distribution of the diboson system 
in the $W^{+}W^{-}$ production process at LHC, $pp$ collisions at 14 TeV.
Together with the summed distribution (solid circles), 
the contribution from the 0-jet (open circles) and the LLL-subtracted 
1-jet (open squares) processes are shown separately.
The two simulations are smoothly combined also in this process.
The GR@PPA result is compared with the results of 
MC@NLO 3.4.2~\cite{Frixione:2008ym} 
and PYTHIA 6.421~\cite{Sjostrand:2006za}.
The MC@NLO result has been obtained with the mode {\tt IL1} = {\tt IL2} = 1.
Thus, we can expect the decay width and spin correlations to be simulated 
properly.
The PYTHIA result has been obtained using its built-in generator 
and the "new" PS model, 
as in the case of the previous results for single weak-boson productions.
The effects of the decay width and spin correlations are simulated 
also in PYTHIA.

We find a reasonable agreement between the GR@PPA and MC@NLO results.
Note that the absolute values of the differential cross section are compared, 
and not the relative shapes.
The overall difference that we can see in the figure can be attributed to the 
lack of non-divergent corrections at the next-to-leading order (NLO) in GR@PPA.
A substantial difference at low $p_{T}$ must be due to the difference 
between PYTHIA and HERWIG, 
employed for small-$Q^{2}$ simulations in GR@PPA and MC@NLO, respectively.
The PYTHIA built-in generator shows a significantly small value of the total 
cross section because it does not include the 1-jet cross section. 
In addition, PYTHIA shows a high-$p_{T}$ behavior apparently different 
from that of GR@PPA and MC@NLO because hard radiations are simulated 
with an extrapolation of the collinear approximation.

\begin{figure}
\begin{center}
\includegraphics[scale=0.65]{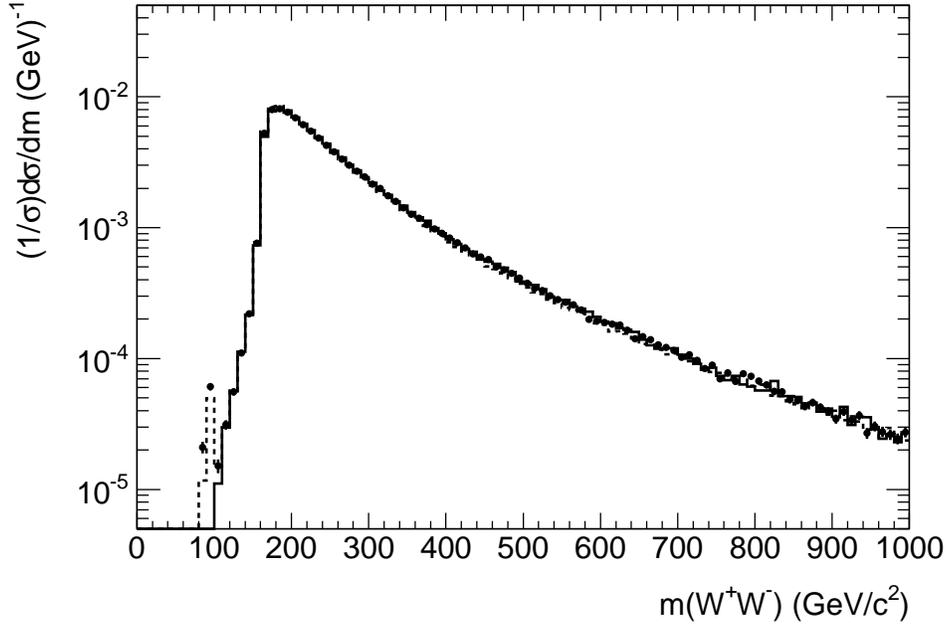}
\caption{\label{fig:ww_mvv}
Invariant mass distribution of the diboson system in the $W^{+}W^{-}$ 
production process at LHC. 
The predictions from GR@PPA (filled circles), MC@NLO (solid histogram), 
and PYTHIA (dashed histogram) are compared.
The results are normalized to the total yield.
}
\end{center}
\end{figure}

Figure~\ref{fig:ww_mvv} shows the relative shape of the invariant mass 
spectrum of the $W^{+}W^{-}$ system.
The results of GR@PPA, MC@NLO, and PYTHIA are in very good agreement with 
each other, except for a small peak below the threshold.
This peak corresponds to the $Z$ resonance 
in which $Z$ decays to a highly virtual $W$ boson pair;
it is visible in the GR@PPA and PYTHIA results, 
whereas it is absent in the MC@NLO result.
The peak vanishes if we require a substantial $p_{T}$ value 
for the decay electrons, {\it e.g.}, $p_{T} >$ 20 GeV/$c$.

\begin{figure}
\begin{center}
\includegraphics[scale=0.65]{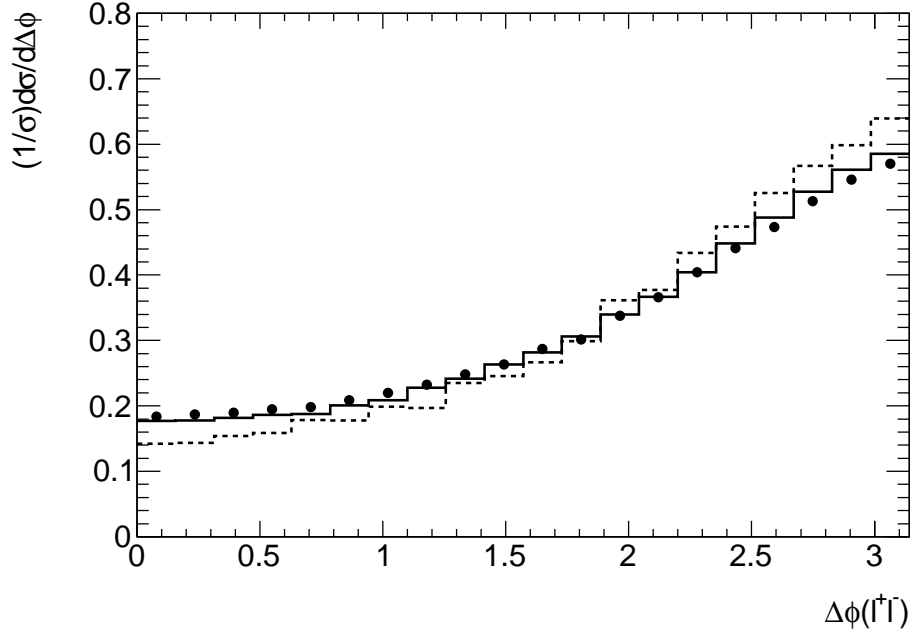}
\caption{\label{fig:ww_dphi_ll}
Relative shape of the azimuthal opening angle ($\Delta\phi$) distribution 
between the two decay electrons from the $W^{+}W^{-}$ pair to be produced 
at LHC via known weak interactions.
The GR@PPA prediction (plot) is compared with the predictions from 
MC@NLO (solid histogram) and PYTHIA (dashed histogram).
}
\end{center}
\end{figure}

The azimuthal correlation between the two decay leptons is frequently used 
as a quantity to distinguish Higgs boson decays to $W^{+}W^{-}$ from 
non-resonant weak interactions that produce them.
The predictions on the opening angle ($\Delta\phi$) distribution from 
GR@PPA, MC@NLO, and PYTHIA are compared in Fig.~\ref{fig:ww_dphi_ll}.
We can see a substantial difference between the three predictions, 
even though the overall tendencies are identical.
This difference reflects the difference in the $p_{T}$ spectrum of the 
diboson system.
Therefore, a detailed understanding/tuning of the $p_{T}$ spectrum will be 
crucial to identify the possible small signal originating 
from the Higgs boson production.

\begin{figure}
\begin{center}
\includegraphics[scale=0.65]{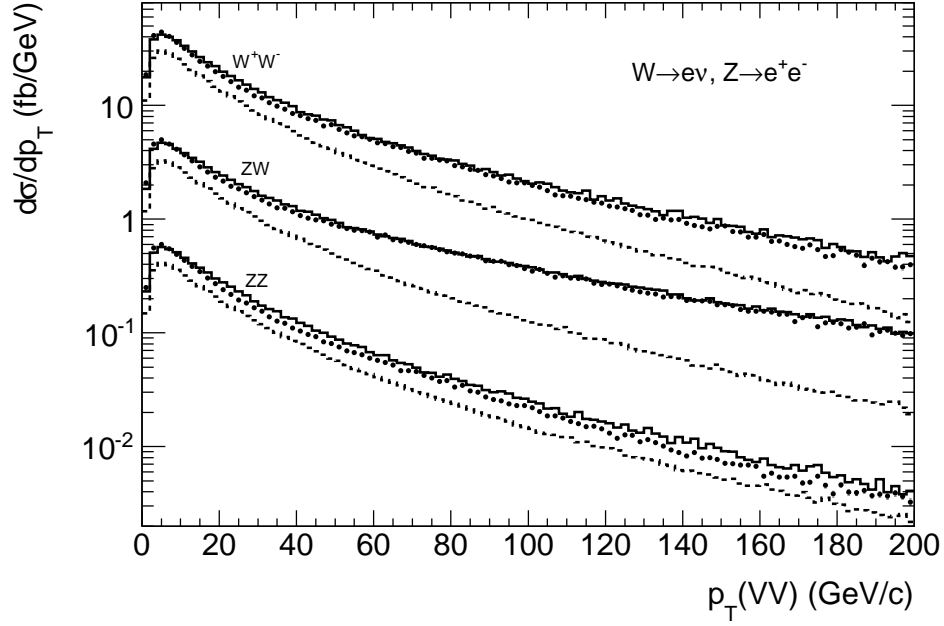}
\caption{\label{fig:ptvv}
$p_{T}$ spectrum of the diboson system for the $W^{+}W^{-}$, $ZW$, 
and $ZZ$ production processes at LHC.
The GR@PPA results (plots) are compared with those of 
MC@NLO (solid histograms) and PYTHIA (dashed histograms).
The MC@NLO simulations were carried out in the mode IL1 = IL2 = 7.
}
\end{center}
\end{figure}

\begin{figure}
\begin{center}
\includegraphics[scale=0.65]{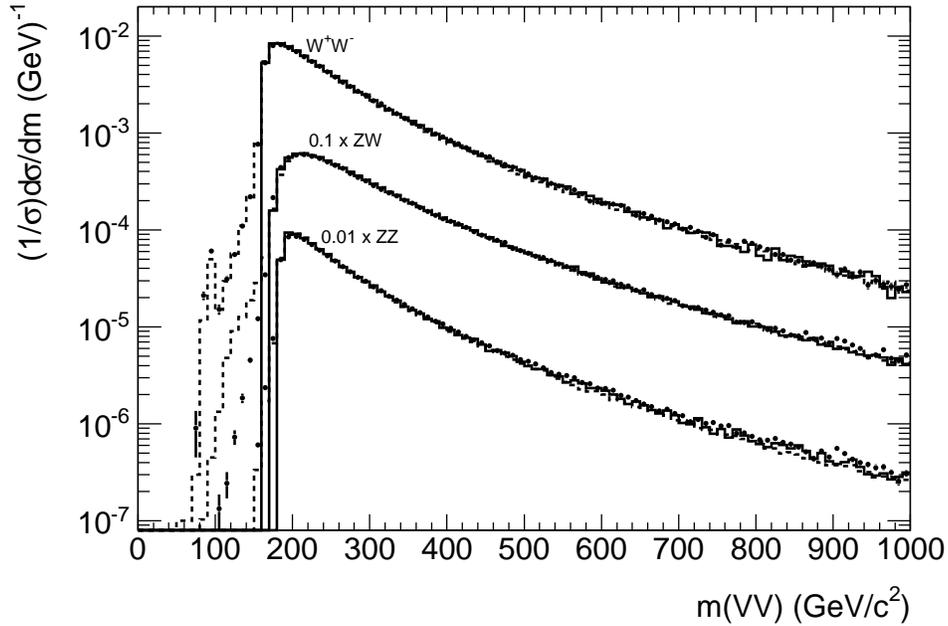}
\caption{\label{fig:mvv}
Relative shape of the diboson invariant mass spectrum 
for the $W^{+}W^{-}$, $ZW$, and $ZZ$ production processes at LHC.
The GR@PPA results (plots) are compared with those of 
MC@NLO (solid histograms) and PYTHIA (dashed histograms).
The MC@NLO simulations were carried out in the mode IL1 = IL2 = 7.
}
\end{center}
\end{figure}

The $p_{T}$ spectrum and the invariant mass spectrum of the diboson system 
are presented for the $W^{+}W^{-}$, 
$ZW$, and $ZZ$ production processes at LHC
in Figs.~\ref{fig:ptvv} and \ref{fig:mvv}, respectively.
The GR@PPA predictions are compared with those from MC@NLO 3.4.2 
and PYTHIA 6.421 in the figures. 
Here, we present the MC@NLO results in the mode of IL1 = IL2 = 7, 
in which MC@NLO produces $W$ and $Z$ bosons as on-shell particles 
having decay widths equal to zero.
The mode IL1 = IL2 = 1 used for the $W^{+}W^{-}$ production is not 
available for the $ZW$ and $ZZ$ productions\footnote{
It has been announced that the spin correlation is included 
in  $ZW$ production since the version 4.0 of MC@NLO.}.
As a result of the zero decay widths, MC@NLO cannot produce diboson pairs 
having invariant masses below the threshold.
The behaviors of the three simulations are similar to those previously 
observed for the $W^{+}W^{-}$ production in the $p_{T}$ spectrum.
At present, we do not identify the reason of a substantial difference 
between GR@PPA and PYTHIA in the $ZW$ mass spectrum below the threshold.

\begin{figure}
\begin{center}
\includegraphics[scale=0.65]{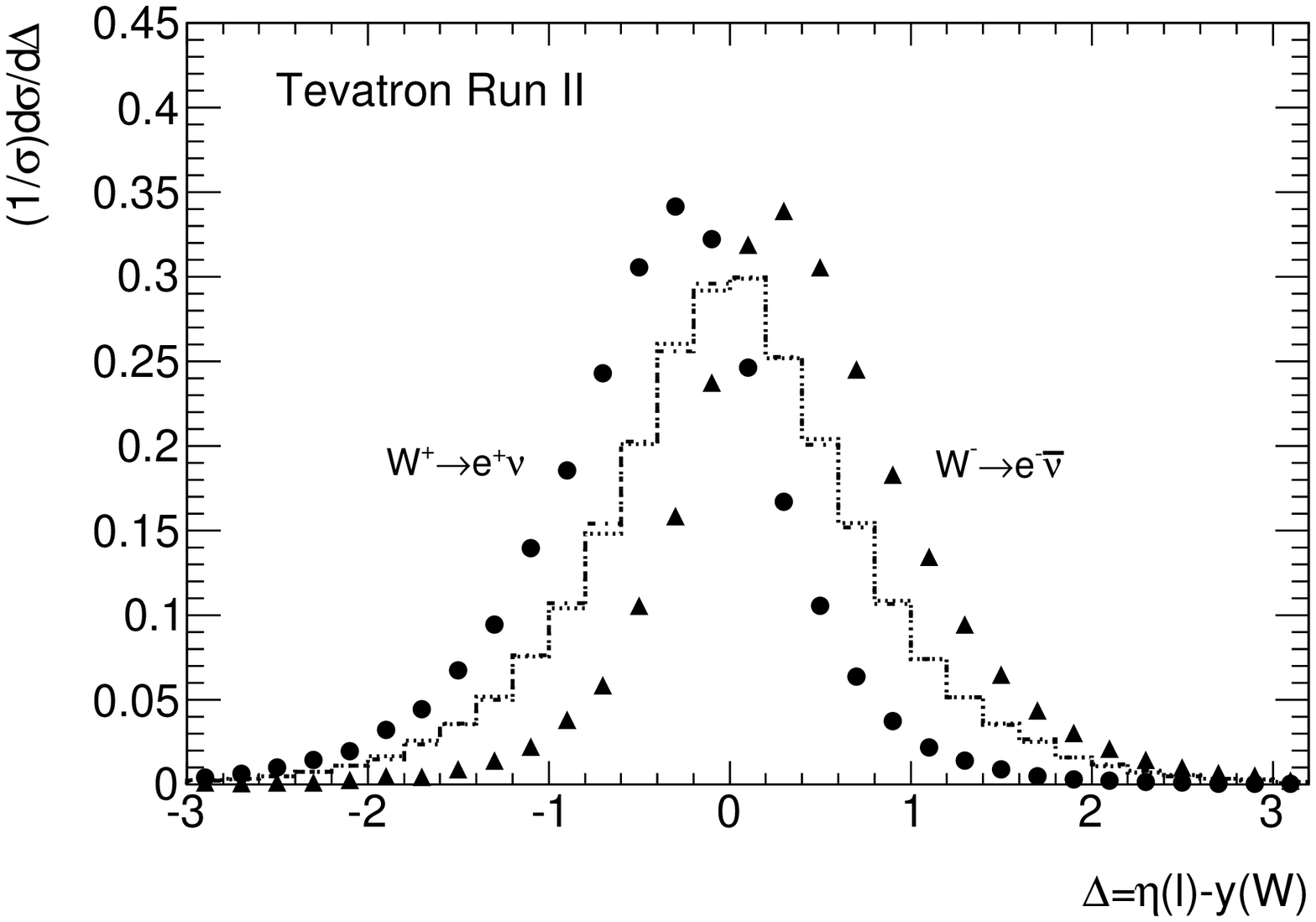}
\caption{\label{fig:zw_tev_deta_lv_mcatnlo}
Simulated distributions of the difference between the pseudorapidity 
of an electron and the rapidity of its parent $W$ boson in the $ZW$ 
production process at Tevatron Run II.
The GR@PPA result (plots) is compared with the MC@NLO result (histograms).
The distributions for $e^{+}$ and $e^{-}$ are shown with filled circles 
and filled triangles, respectively, for the GR@PPA result, 
and with dotted and dot-dashed histograms, respectively, for the MC@NLO result.
}
\end{center}
\end{figure}

As a result of the on-shell approximation, the spin information is not 
propagated to the decay kinematics in these MC@NLO simulations.
Although this effect is small in observable quantities, 
it may affect detailed studies on the production mechanism.
As an example, in Fig.~\ref{fig:zw_tev_deta_lv_mcatnlo}, we show
the distribution of the difference between the pseudorapidity of the electron 
and the rapidity of its parent $W$ boson in the $ZW$ production process 
at Tevatron.
We can see an apparent difference between the distributions for $e^{+}$ 
and $e^{-}$ in the GR@PPA simulation result, 
whereas there is no such difference in the MC@NLO simulation. 
Though it is not shown here, 
the asymmetry observed in the GR@PPA simulation is in good agreement with 
the result of the PYTHIA simulation.

\section{Practical performance}
\label{sec:perf}

The portability of the program has been tested on some recent Linux systems: 
Scientific Linux CERN (SLC) 4 and 5, Fedora 10 and 13, and Ubuntu 9.
We have successfully executed the sample programs on these systems.
Hence, the compilers that we have tested are gcc version 3 and 4 up to 4.4.
The Fortran compilers are g77 in gcc version 3 and gfortran in gcc version 4.
The difficulties that we have encountered are mostly 
related to the external libraries.
Although there is no difficulty in the use of g77, 
we need to add the {\tt -fsecond-underscore} option to the Fortran compilation 
when we use RBOOK/ROOT together with gfortran.
In addition, some codes in PYTHIA are incompatible with the simultaneous 
use of RBOOK/ROOT and gfortran.
Necessary modifications for the use of gfortran are described in the file 
named {\tt config.input-gcc4} in the GR@PPA 2.8 package.
In general, we have to be careful about the incompatibility between 
minor versions of gcc4.
It is better for users to compile external libraries by themselves 
in the same environment as GR@PPA, 
unless the libraries are provided as standard packages of the distribution.
Though this is independent of GR@PPA, 
we need to replace {\tt /bin/sh} in the setup scripts of LHAPDF with 
{\tt /bin/bash} when we install it on Ubuntu 
because {\tt /bin/sh} is a symbolic link to {\tt /bin/dash}.
In addition, when we install LHAPDF on a computer having relatively 
small memory, it is better to add the {\tt --enable-low-memory} option 
in the configuration because LHAPDF requires a memory larger than 1 GB 
if it is installed without this option.

The practical performance of GR@PPA 2.8 was intensively tested 
on the SLC4 operating system installed on an Intel Xoen 5160 CPU 
operated at a clock speed of 3.0 GHz.
The compiler was gcc version 3.4, {\it i.e.}, g77 for Fortran.
Though the operating system was 64 bit, all the tests were carried out  
with the 32-bit mode, 
{\it i.e.}, the {\tt -m32} option was used in the compilation.
The used external libraries are LHAPDF 5.7.0, CERNLIB 2006, 
and ROOT 5.22.00.
CERNLIB was installed from a standard package repository, 
and the binary package of ROOT was downloaded from the official Web page.
LHAPDF was locally compiled from the source distribution.

The size of the executable module, {\tt grappa}, is 1.6 MB 
for the sample programs using HBOOK, and 1.2 MB for those using RBOOK, 
when all the processes are retained.
The memory size required for the execution is about 6.6 MB with HBOOK 
and 30 MB with RBOOK when the built-in CTEQ6 PDF is used.
If LHAPDF is used instead, the memory size is increased to 1.5 GB 
in both cases.
However, if we use LHAPDF built with the {\tt --enable-low-memory} option, 
the size is reduced to 150 MB and 170 MB, respectively.
The program size can be reduced by removing unnecessary processes.
Actually, if we retain only the {\tt wjets} directory, 
the total size of the package is reduced from the original size of 5.1 MB 
to 3.0 MB.
The size of {\tt grappa} is also reduced to 0.96 MB using HBOOK 
and to 0.57 MB using RBOOK.
However, the reduction in the memory size is small, at most 10\%,
even when the built-in CTEQ6 is used for the PDF.

\begin{table}[t]
\caption{CPU time consumed by GR@PPA 2.8 on a 3.0-GHz Intel Xeon processor 
with the SLC4 operating system.
The time required for the cross-section integration by BASES 
and that for the event generation by SPRING are separately presented.
The former is in the unit of minutes and the latter is in seconds 
for generating every 1k events.
Results are presented for two choices of parton showers (PS).
If QCDPS is used for simulating the initial-state parton shower, 
it is applied in the second stage of BASES for the integration, 
while no PS is applied during the integration if QCDPSb is chosen.}
\label{tab:CPU}
\begin{center}
\begin{tabular}{c c c c c}
\hline
& \multicolumn{2}{c}{QCDPS} &  \multicolumn{2}{c}{QCDPSb}\\
& integ (min) & gen (sec/1k) & integ (min) & gen (sec/1k)\\
\hline
$W$ & 9.9 & 23 & 0.72 & 4.2 \\
$Z$ & 9.3 & 18 & 0.73 & 4.2 \\
\hline
$W^{+}W^{-}$ & 350 & 260 & 120 & 27 \\
$ZW$ & 200 & 190 & 75 & 16 \\
$ZZ$ & 220 & 96 & 89 & 12 \\
\hline
\end{tabular}
\end{center}
\end{table}

The CPU time consumed for the execution of GR@PPA 2.8 is summarized in 
Table~\ref{tab:CPU}.
The program was executed with the default settings in the sample programs 
for the LHC condition in the results denoted by QCDPS.
The results are separately presented for the integration by BASES 
and the event generation by SPRING.
In general, the BASES integration time is dominated by that for the 1-jet 
processes, and by the second stage where QCDPS is applied.
The time consumption when QCDPSb ({\tt ishower} = 4) is used instead of 
QCDPS ({\tt ishower} = 3) is also presented in Table~\ref{tab:CPU}, 
in which parton showers are not applied in the integration;  
instead, PDF is used to derive the parton distribution at large $Q^{2}$.
The integration time is significantly reduced by this change, as expected.
The event generation time is predominantly determined by that for the 0-jet 
processes because the 1-jet cross section is far smaller.
The generation time is presented in seconds consumed 
for the generation of 1k events.
The actual running time increases linearly according to the number of events 
that users require.

\section{Summary}
\label{sec:sum}

We have described a new release of the GR@PPA event generator package, 
GR@PPA 2.8, for proton-proton and proton-antiproton collisions.
This release supports an initial-state jet matching method, 
the limited leading-log (LLL) subtraction, that we have proposed.
The matching method can be applied to single $W$ and $Z$ production processes 
and diboson ($W^{+}W^{-}$, $ZW$ and $ZZ$) production processes.
Custom-made parton shower (PS) programs are included in the package 
in order to ensure satisfactory performance of the matching method.
Though the used matrix elements remain at the tree level, 
we can reproduce the recoil effects of QCD radiations in the entire phase space 
of the weak-boson system by combining the "0-jet" and "1-jet" processes 
with the help of the matching method. 
The decay widths of weak bosons and the spin effects in the decay products 
are exactly simulated at the tree level 
because the decays are included in the matrix elements, 
as in the previous versions.

The event generators support the use of the LHAPDF library for evaluating 
the parton distribution function (PDF) inside the proton and antiproton.
In addition, a built-in PDF is provided for tests. 
The classical PDFLIB library can also be used.
Though we can only use leading-order PDFs when a forward-evolution PS 
for the initial state (QCDPS) is chosen,
no such restriction exists if we choose a backward-evolution PS (QCDPSb).
A PS for the final state (QCDPSf) based on a new concept is also provided 
for completeness.
The generated events are stored in a file using the LhaExt utility, 
a set of simple I/O routines for the LHA user-process interface.
The event records written in the file can be fed to PYTHIA 
in order to perform simulations down to the hadron level.
The parton showers in GR@PPA can be turned off 
if users want to apply them in external libraries.
The "new" PS model in PYTHIA exhibits a performance nearly identical to 
QCDPS and QCDPSb.

The program is distributed as an all-in-one package, including 
process-dependent routines together with the event generator framework.
Sample programs are provided for each process.
The installation and execution of sample programs are easy 
in UNIX/Linux systems.
The installation of the GR@PPA libraries can be completed in one action.
Users are allowed to remove processes in which they are not interested 
before the installation.
The procedure for building sample programs consists of two steps: 
a configuration and the compile/link.
Once it is configured, the sample programs become relocatable.
Users can copy the program directory to any place
in order to customize it for their own studies.

We provide two sample programs for each process.
The difference between the two programs is in the histogramming tool only.
HBOOK in CERNLIB is used in one of the two programs, 
while RBOOK, a simple Fortran interface to ROOT, is used in the other.
Therefore, in order to test the sample programs, 
users have to prepare the CERNLIB library or the ROOT library 
in addition to PYTHIA and LHAPDF libraries according to their preference.
The produced histogram files can be manipulated by PAW in the former 
and by ROOT in the latter.

The performance of the event generators in GR@PPA 2.8 has been tested 
by comparing the results with those of PYTHIA and MC@NLO.
In general, the results are in excellent agreement.
Most of the observed differences can be understood as the effects of 
different implementations and approximations in these programs. 
We can, at least, conclude that there is no apparent mistake 
in GR@PPA 2.8 and the other programs that we have compared.

The portability of GR@PPA 2.8 has been tested on several recent Linux systems. 
Problems which users may encounter are mostly related to compatibility 
with external libraries.
Users need to be careful about the consistency in the compiler version 
and compile options.
The program size and execution time have also been studied extensively.
The program size is predominantly determined by external libraries 
which users select, 
while the execution time is nearly independent of them.
The application of QCDPS consumes much CPU time, not only in integration 
but also in event generation. 
The use of QCDPSb reduces them, as expected.

\section*{Acknowledgments}

This work has been carried out as an activity of the NLO Working Group, 
a collaboration between the Japanese ATLAS group and the numerical analysis 
group (Minami-Tateya group) at KEK.
The authors wish to acknowledge useful discussions with the members, 
especially S. Tsuno and J. Fujimoto of KEK, and K. Kato of Kogakuin U.












\end{document}